\documentclass[twocolumn,times,svgnames]{aastex701}

\usepackage{multirow}
\usepackage{mathtools}
\usepackage{graphicx}

\colorlet{hlwas-wide-color}{PaleGreen}
\colorlet{hlwas-medium-color}{LimeGreen}
\colorlet{hlwas-deep-color}{DarkOliveGreen}
\colorlet{elais-color}{MediumBlue}
\colorlet{edfs-color}{RoyalBlue}
\colorlet{gps-color}{FireBrick}
\colorlet{gbtds-color}{Gold}

\renewcommand{\descriptionlabel}[1]{\hspace{\labelsep}\texttt{#1}}

\graphicspath{{./}{figures/}}

\received{}
\revised{}
\accepted{}

\shorttitle{The near field with Roman}
\shortauthors{Sanderson et al.}

\newcommand{\K}{\,\textnormal{K}}

\newcommand{\upenn}{Department of Physics \& Astronomy, University of Pennsylvania, Philadelphia, PA 19104, USA}
\newcommand{\columbia}{Department of Astronomy, Columbia University, 550 West 120th Street, New York, NY 10027, USA}
\newcommand{\uw}{Dept of Astronomy, Box 351980, University of Washington, Seattle, WA 98115}
\newcommand{\stsci}{Space Telescope Science Institute, 3700 San Martin Drive, Baltimore, MD 21218, USA}
\newcommand{\rutgers}{Department of Physics and Astronomy, Rutgers, the State University of New Jersey, 136 Frelinghuysen Road, Piscataway, NJ 08854, USA}
\newcommand{\utoronto}{David A. Dunlap Department of Astronomy \& Astrophysics, University of Toronto, 50 St George Street, Toronto, ON M5S 3H4, Canada}
\newcommand{\uva}{Department of Astronomy, University of Virginia, 530 McCormick Road, Charlottesville, VA 22904 USA}
\newcommand{\dark}{DARK, Niels Bohr Institute, Jagtvej 155A, 2200 Copenhagen N, Denmark}
\newcommand{\berkeley}{Department of Astronomy, University of California, Berkeley, CA 94720-3411, USA}

\begin{document}

\title{A near field guide to Roman's wide-area surveys}

\correspondingauthor{Robyn Sanderson}
\email{robynes@sas.upenn.edu}
\author[0000-0003-3939-3297]{Robyn E. Sanderson}
\affiliation{\upenn}
\email{robynes@sas.upenn.edu}
\author[0000-0001-7494-5910]{Kevin A. McKinnon}
\affiliation{\utoronto}
\email{kevin.mckinnon@utoronto.ca}
\author[0000-0001-7928-1973]{Adrien C.R. Thob}
\affiliation{\upenn}
\email{athob@sas.upenn.edu}
\author[0000-0002-7502-0597]{Benjamin Williams}
\affiliation{\uw}
\email{benw1@uw.edu}
\author[0000-0001-6584-6144]{Kiyan Tavangar}
\affiliation{\columbia}
\email{kt2587@columbia.edu}
\author[0000-0002-6021-8760]{Andrew B. Pace}
\affiliation{\uva}
\email{apace@virginia.edu}
\author[0000-0001-8738-6011]{Saurabh W. Jha}
\affiliation{\rutgers}
\email{saurabh@physics.rutgers.edu}
\author[0000-0003-3136-9532]{Javier S\'{a}nchez}
\affiliation{\stsci}
\email{jsanchez@stsci.edu}
\author[0000-0002-5865-0220]{Abigail Lee}
\affiliation{\berkeley}
\email{abby.lee@berkeley.edu}
\author[0000-0003-0256-5446]{Sarah Pearson}
\affiliation{\dark}
\email{sarah.pearson@nbi.ku.dk}

\begin{abstract}
    The Nancy Grace Roman Space Telescope currently plans to survey nearly 6000 square degrees of the sky, mainly in the High-Latitude Wide-Area Survey (HLWAS) and Galactic Plane Survey (GPS). Although these surveys are optimized for other science, they are also a treasure trove for studying the nearby universe. The foreground of the HLWAS includes 59 known stellar streams, 14 known satellite galaxies, and 9 globular clusters in the Milky Way, and an additional 63 galaxies within 10 Mpc spanning several orders of magnitude in stellar mass. The GPS includes an additional 38 globular clusters in its footprint. We summarize and visualize these populations and discuss some of the relevant characteristics of the planned Roman observations. We also examine the expected astrometric performance of the core surveys based on the anticipated time-baselines between observations, and point out the substantial improvement provided by longer time intervals between repeat observations. In particular, the plan for a 6-month revisit timescale in the HLWAS is a missed opportunity from the perspective of proper motions. These data will nonetheless be a powerful new resource for studying the Milky Way and its neighborhood. To aid in planning, the data and code used to generate the figures and tables in this paper are available publicly at \url{github.com/Dynamics-Penn/roman-nearfield}.
\end{abstract}


\section{Overview}

The Nancy Grace Roman Space Telescope (Roman) currently plans to survey nearly 6000 square degrees of the sky \citep{2025arXiv250510574O} in several main components. The High-Latitude Wide-Area Survey (HLWAS), which includes deep, medium and wide tiers, will cover 2415 square degrees in three filters and an additional 2700 square degrees in a single filter. The High-Latitude Time-Domain Survey (HLTDS) will observe several small fields within this area repeatedly at a range of cadences. The Galactic Plane Survey (GPS) will observe an additional $\sim$700 square degrees starting early in the mission \citep{2025arXiv251107494G}.  The prominent characteristics of these surveys are reiterated in Table \ref{tbl:survey}. 

The implementation of the HLWAS is driven by the requirements for achieving the desired performance of Roman's core program for cosmology. However, many of its characteristics are also close to ideal for observing near-field targets and for obtaining proper motions in revisited fields, given the exquisite control over the point-spread function (PSF) that is required to meet weak-lensing goals (Section \ref{sec:roman_PMs}). The 59 streams (Section \ref{sec:streams}) and 77 nearby galaxies (Section \ref{sec:nbgs}) in the foreground of the HLWAS and the 47 globular clusters (Section \ref{sec:gcs}) in the GPS and HLWAS combined will all benefit from these requirements. For near-field cosmologists, these observations will be a treasure trove of information about the Milky Way and Local Volume. Here we attempt to summarize the near field in the HLWAS footprint, and highlight some implications for the near field of the design choices of the survey.

\begin{table*}[htp]
\begin{center}
\caption{Characteristics of the Roman High Latitude Wide Area Survey (HLWAS), Roman High Latitude Time Domain Survey (HLTDS), and Roman Galactic Plane Survey (GPS). Complete information is available in \citet{2025arXiv250510574O} for the HLWAS and HLTDS and \citet{2025arXiv251107494G} for the GPS. }
\begin{tabular}{|p{1in}ccp{1.25in}c|}
\hline
Component & Area (deg${}^2$) & Filters & Point-source Depth, 5$\sigma$ AB mag & Color \\
\hline
HLWAS Wide & 2700 & H & 26.2 & \colorbox{hlwas-wide-color} {\phantom{block}} \\
HLWAS Medium & 2415 & YJH+grism &  26.5 &  \colorbox{hlwas-medium-color} {\phantom{block}}\\
HLWAS Deep & 19.2 & ZYJHFKW+grism & 25.9 (K)--28.3 (W) & \colorbox{hlwas-deep-color} {\phantom{block}} \\
\hline

HLTDS North Wide (ELAIS) & 10.68 & RZYJH & \multirow{2}{1.25in}{10-day cadence, 60-300 sec exposures, alternating filters. } & \colorbox{elais-color} {\phantom{block}} \\
HLTDS South Wide (Euclid S) & 7.59 & RZYJH &  & \colorbox{edfs-color} {\phantom{block}} \\
\hline
GPS & 691 & JHFK & 23--24 & \colorbox{gps-color}{\phantom{block}} \\
GBTDS & \multicolumn{3}{l}{Included for completeness} & \colorbox{gbtds-color}{\phantom{block}} \\
\hline

\end{tabular}
\end{center}
\tablecomments{\textit{Filters:} R=F062, Z=F087, Y=F106, J=F129, H=F158, F=F184, K=F213, W=F146 broad-band. \textit{Color:} the color used to mark the footprint of each component in the maps in this paper.}
\label{tbl:survey}
\end{table*}%

\section{Astrometry with Roman} \label{sec:roman_PMs}

\begin{figure*}[ht]
    \centering
    \includegraphics[width=\linewidth]{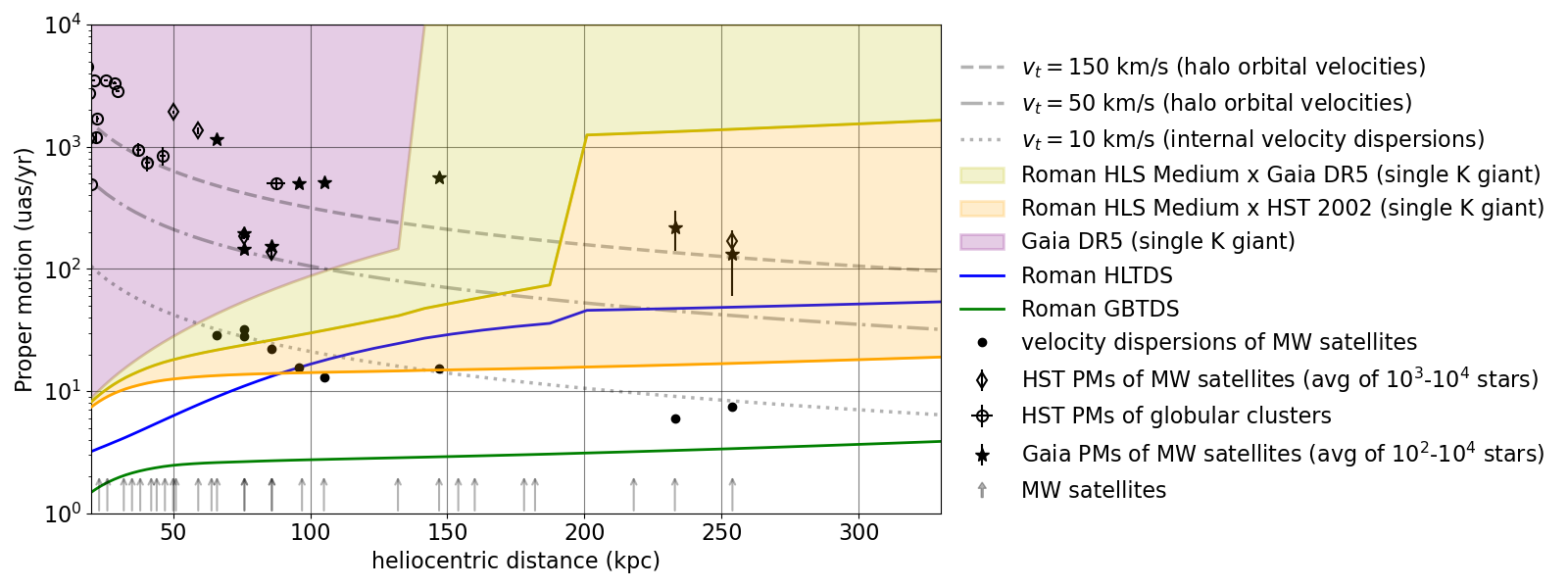}
    \includegraphics[width=\linewidth]{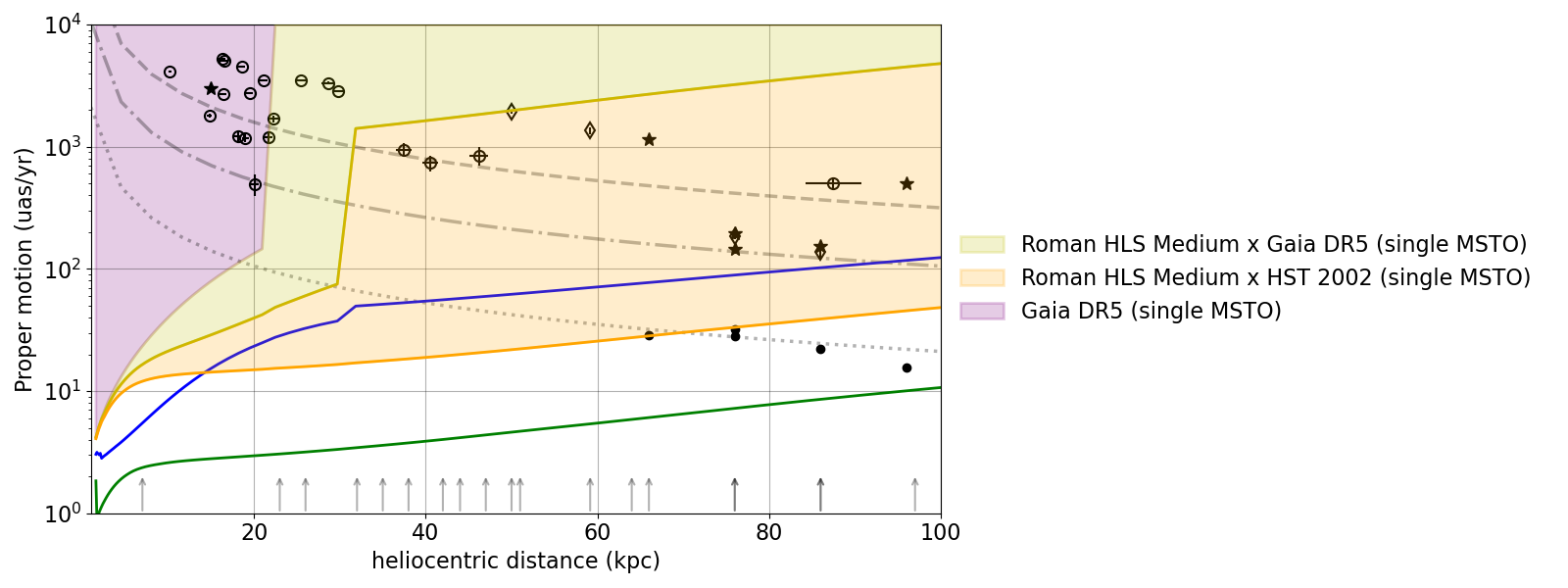}
    \caption{The reach of Roman proper motions. Magnitudes have been converted to distances assuming K giants with absolute magnitude $G=2$, color $B_P-R_P = 1.5$ (top) and MSTO stars with absolute magnitude $G=4.5$, color $B_P-R_P = 1$ (bottom). As noted in the legend, dotted and dashed curves show typical proper motions for orbital and internal velocities; vertical arrows denote locations of satellites; points denote measured velocities in the MW. The pink shaded region is the reach of Gaia DR5. Yellow is the additional reach of combining the HLWAS with Gaia, assuming a final epoch with a 5 year baseline. Orange shows the additional gain possible for fields with HST coverage, which can date back as far as 2002. Solid green and blue lines estimate PM precisions for the GBTDS and HLTDS, respectively.}
    \label{fig:pms_dist}
\end{figure*}

Roman's wide-field surveys offer the potential for significant gain over Gaia DR5 \citep{2016A&A...595A...1G} in the reach of accurate proper motion (PM) measurements in the near field (Figure \ref{fig:pms_dist}), especially for stars with $G>21.5$~mag, those that are heavily extincted at optical wavelengths, and those that are intrinsically red. In addition, giant stars observed by Gaia are intrinsically about 0.5 mag brighter in the Roman bands (based on Parsec isochrones; \citealt{bressan12, Tang2014}), further increasing Roman's range relative to Gaia. With the right choices in scheduling repeat observations of the same field, we will be able to obtain single-star PMs at similar precision to Gaia for a factor of two increase in distance. Targets with archival HST observations can extend this region nearly to the MW's virial radius and provide single-star access to internal velocities out to $\sim$100 kpc from the Sun. These improvements will take our ability to study the dynamics of our Galaxy into a new era.
\begin{figure*}[t]
    \centering
    \includegraphics[trim={0cm 1cm 0.5cm 2.75cm},clip,width=1.0\linewidth]{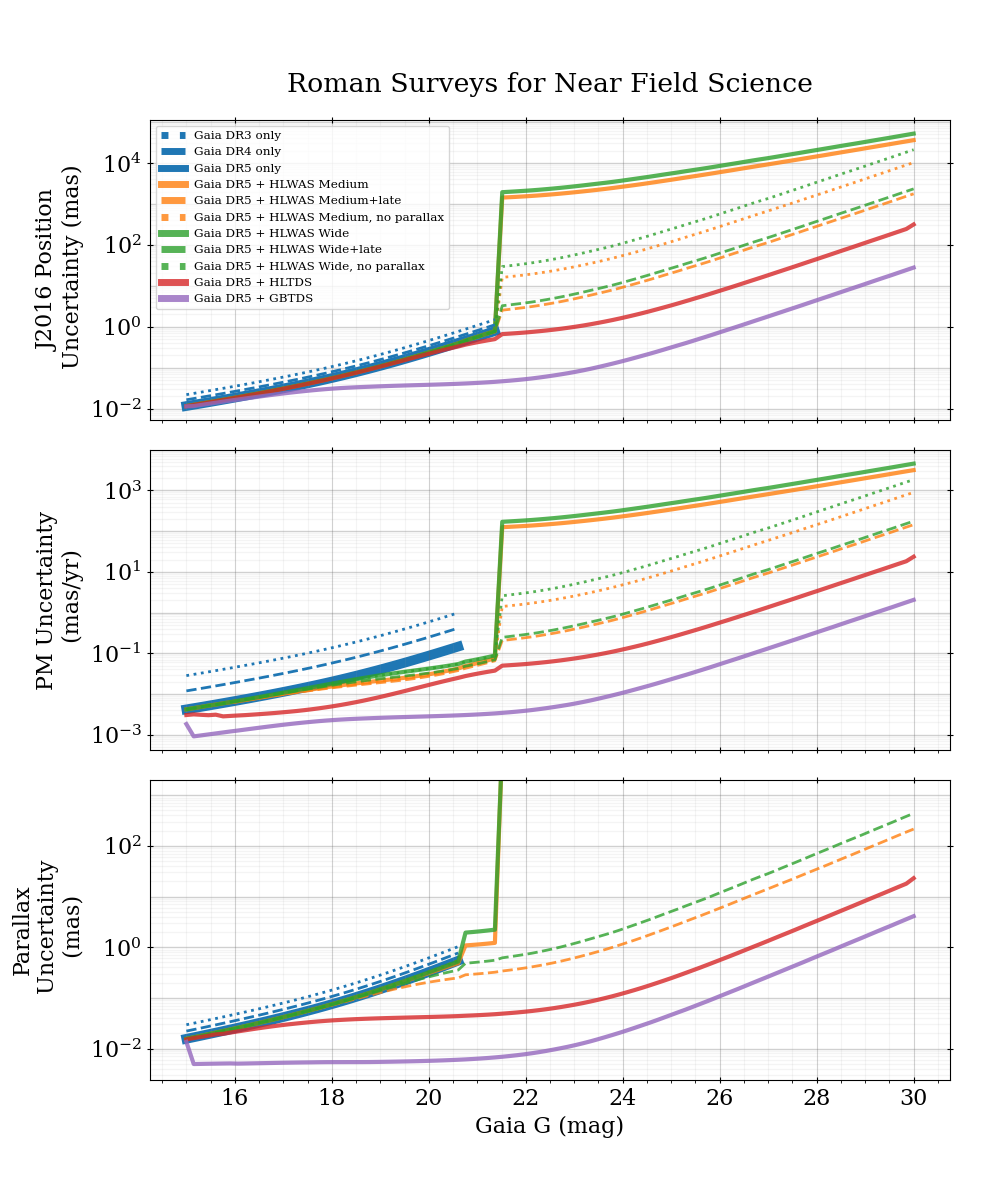}
    \caption{Expected astrometric precision as a function of magnitude for different Roman surveys relevant to Near Field science. The precise configurations (e.g., exposure time, timing of epochs, pointing) used for these simulations are described in the text and \citet{McKinnon_2026}. Here, we use the color information of an old ($12.7$~Gyr) main-sequence turnoff (MSTO) star when incorporating information from different Gaia and Roman filters. Gaia DR3-DR5 astrometric precisions are shown by the different blue lines, which extend to 21.5~mag in position and 20.7~mag in parallax and proper motion. Position uncertainties are given with respect to the Gaia reference epoch of J2016.0. The GBTDS and HLTDS are both expected to produce high quality parallaxes and PMs. The HLWAS, when observed with a 6 month time baseline using 2 epochs early in Roman's life (solid lines) do not yield scientifically useful PMs or parallaxes for faint stars ($G>21.5$~mag). Adding a late HLWAS epoch 5 years after the first (dashed lines) dramatically improves the astrometric precision by more than 3 order of magnitude for $G=21.5$~mag. When apriori distances are available such that parallaxes can be independently constrained, proper motion uncertainties for the 6 month baseline HLWAS can be improved (dotted lines), but this option will not be available for the majority of stars.}
    \label{fig:Roman_astrometry_errs}
\end{figure*}

\begin{figure*}[t]
    \centering
    \includegraphics[trim={0cm 0cm 0cm 1.5cm},clip,width=1.0\linewidth]{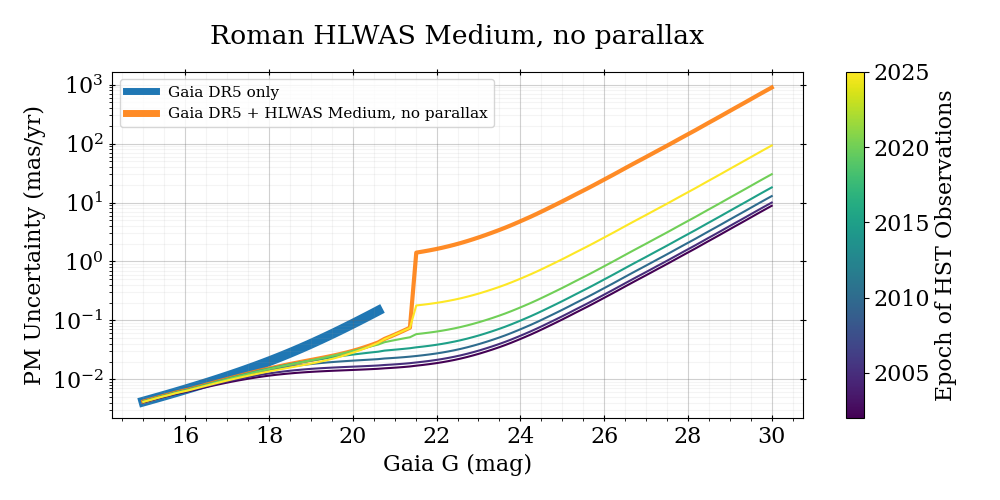}
    \caption{Expected proper motion precision as a function of magnitude for an old MSTO star in the HLWAS Medium tier (i.e. dotted orange line in middle panel of Figure~\ref{fig:Roman_astrometry_errs}) with a single additional epoch of HST (colored lines). As expected, an early HST epoch near 2002 leads to the largest time baseline with Roman, producing the most precise proper motions. We note that only a small-but-powerful fraction of the sky has access to these archival HST images.}
    \label{fig:HLWAS_Medium+HST_astrometry_errs}
\end{figure*}

 However, realizing this potential requires some care in the choice of epochs. In this section we discuss the implications of various different options for scheduling repeat visits to the same field in the HLWAS, which has the most scheduling flexibility of any of the core surveys since its core science, weak lensing, does not rely on time-domain data. We also show predictions for the High-Latitude Time Domain Survey (HLTDS) and Galactic Bulge Time Domain Survey (GBTDS), which have specific core science related to the time domain, using the most recent published strategy as of the publication of this paper \footnote{available at \url{https://roman-docs.stsci.edu}}. 
 
 For all the scenarios we present, the expected astrometric precision from Roman is calculated \href{https://github.com/KevinMcK95/gaia_roman_astrometry}{using this publicly available tool}\footnote{a helpful tutorial for using the Gaia+Roman astrometry tool is described on this \href{https://roman.ipac.caltech.edu/page/roman-gaia-astrometry-tool}{ Roman Science Support Center at IPAC} webpage}. The exact method for the astrometry calculations---inspired by the Gaia+HST tool \texttt{BP3M} \citep{McKinnon_2024}---is described in detail in \citet{McKinnon_2026}. In brief, this tool generates realistic estimates of Roman position uncertainty as a function of magnitude and exposure time using \texttt{stpsf}\footnote{see \texttt{stpsf} description at \url{https://stpsf.readthedocs.io/en/latest/intro.html}} \citep{Perrin_2012,Perrin_2025} and \texttt{pandeia}\footnote{see \texttt{pandeia} description at \url{https://roman-docs.stsci.edu/simulation-tools-handbook-home/roman-wfi-exposure-time-calculator/pandeia-for-roman/overview-of-pandeia}} \citep{Pontoppidan_2016}, and then combines that information with expected Gaia astrometry precisions\footnote{see the Gaia astrometry versus magnitude relationships presented here \url{https://www.cosmos.esa.int/web/gaia/science-performance}} to derive Gaia+Roman position, parallax, and proper motion (PM) uncertainties. We note that Gaia positions extend to $G=21.5$~mag, while the parallaxes and PMs only extend to $G=20.7$~mag. This means that $G>21.5$~mag stars will only be using Roman information in their astrometry calculations. For $20.7<G<21.5$~mag sources, the Gaia positions serve as an additional epoch at J2016.0. For $G<20.7$~mag, Gaia provides prior knowledge about the full 5D astrometric solution (2D position, 2D PM, 1D parallax). For all of the calculations presented here, we will combine Gaia DR5 expectations with Roman images. The specific filters, exposure times, and observation cadence used in our calculations are those defined on the \href{https://roman-docs.stsci.edu/roman-community-defined-surveys}{Roman Community Defined Surveys} webpage. Thus, given a ($\alpha,\delta$) pointing and a set of Roman epochs, exposure times, filters, and magnitudes, we can calculate the resulting Gaia+Roman combined astrometry. 

Due to their relevance for near-field science, we focus on the HLWAS, HLTDS, and GBTDS. \citet{McKinnon_2026} describes how these predictions were generated and also presents results for the Galactic Plane Survey (GPS). To briefly summarize, the HLTDS and GBTDS will both use a large number of Roman epochs spread over many years, so we expect to measure high precision PMs and parallaxes for both of these surveys. The exact cadence and starting time of the HLWAS observations are not yet settled, however, so we test different configurations for their impact on the astrometric precision. To be able to include parallax effects in our calculations, we must chose an RA, Dec pointing to target. For the following results, the HLWAS simulations point at the Medium Field 1, the HLTDS simulation points at the Euclid Deep Field, and the GBTDS simulation points near the Galactic Center.

Current plans\footnote{See \url{https://roman-docs.stsci.edu/roman-community-defined-surveys/roman-observations-in-the-first-two-years-of-science-operations}} suggest that most of the observations for the deep HLWAS tier will be completed within the first year of the survey, and about half of the Medium tier will be completed in the first two years. Paired with the Community Defined Survey information, this means that we expect to have 2 epochs, separated by 6 months to vary the roll angle, taken early in Roman's lifetime (near J2027.0). The Wide tier is defined to use only the H/F158 filter while the Medium tier is set to use the H/F158, J/F129, and Y/F106 filters. If the HLWAS does indeed use this short time baseline, we do not expect to measure high precision PM or parallax constraints. We supplement the simulations of the 6-month baseline HLWAS with an additional Roman observation 5 years after the first exposure to highlight the impact that a final epoch can have. 

The results of our astrometry simulations are shown in Figure~\ref{fig:Roman_astrometry_errs}, with blue lines showing different Gaia data releases for comparison. As expected, the HLTDS and GBTDS both expect to yield very precise parallaxes and PMs out to very faint magnitude sources. The extreme number of GBTDS exposures ($>54,000$) over a full 5 year period lead to the smallest uncertainties of any survey. The solid HLWAS lines show the results of using a 6 month time baseline with the first epoch occurring in the first half of 2027. Here, we see that the parallaxes will be effectively unconstrained and the PM uncertainties will be very large ($>100$~mas/yr) for faint stars ($G>21.5$~mag), far below the required velocity precision for most near-field science questions. 

While 6-month revisits to the survey fields are practical for the core science of the HLWAS, they fail to harness the incredible astrometric precision of Roman imaging. If this schedule holds, a final epoch near the end of the nominal 5 year mission lifetime would dramatically increase its impact on Milky Way and Local Group dynamical science. The orange and green dashed lines in Figure \ref{fig:Roman_astrometry_errs} show the results of adding a single followup observation with H/F158 exactly 5 years after the first epoch. This final epoch \textbf{improves parallax and PM precision by more than a factor of 1000} for $G=21.5$~mag stars. Using this strategy, we will be able to measure Gaia-DR3 quality parallaxes out to $G\approx22.5$~mag and PMs out to $G\approx24$~mag, effectively extending our dynamical view of the Milky Way to the edge of the galactic halo (Figure \ref{fig:pms_dist}).

For some stars, we will not necessarily need to simultaneously fit the parallaxes and PMs together because we will have strong literature constraints; for example, when focusing on known member stars of well-studied globular clusters and dwarf galaxies. In this case, the PM precision can be improved because degeneracies with parallax are broken. We show the effect of turning off parallax fitting for the HLWAS 6 month baseline case (i.e., solid orange and green lines) with the dotted orange and green lines. Now, PMs are $\sim 100$ times more precise at $G=21.5$~mag compared the HLWAS 6 month strategy. However, we emphasize that this process will only apply to stars that can confidently be assigned a high-precision, apriori distance measurement. This will likely not apply to many foreground stars in the MW stellar halo, which will also potentially lead to difficulties in assigning membership probabilities to different kinematic structures. 


For some fortunate cases, we have access to archival imaging of Near Field structures (e.g. dwarf galaxies, globular clusters, stellar streams) from other telescopes, such as HST. We simulate the effect of adding a single epoch of HST images to the HLWAS Medium tier in Figure~\ref{fig:HLWAS_Medium+HST_astrometry_errs}. We use the no-parallax-fitting version of our HLWAS+Gaia results because many of these structures have well-measured distances. For the HST observations we assume 4 dithers at a given epoch, with 0.5~mas (i.e., 1\% of an ACS/WFC pixel) position uncertainty per image and no magnitude dependence. To be clear, the 2002 results do not also include adding the 2025 results; in every case, we show the orange line combined with a single HST epoch to produce the colored lines. As expected, the PMs significantly improve when we have access to a large time baseline between Roman and HST for $G>21.5$~mag stars. However, we must stress that only a very small fraction of the sky has access to such archival HST images. Even around Near Field structures, HST images often only cover a small range of, for example, galactocentric radii. Using it's large field of view, Roman will be able to tile an area approximately 100 times larger than a single HST pointing. One of Roman's most powerful abilities will come from its ability to provide deep, wide, and precise astrometry, so surveys like HLWAS benefit most from optimizing their observing strategies, not relying on serendipitous archival observations from other telescopes. 

\section{Milky Way Streams}
\label{sec:streams}

\begin{figure*}[htbp]
\begin{center}
\includegraphics[width=\textwidth]{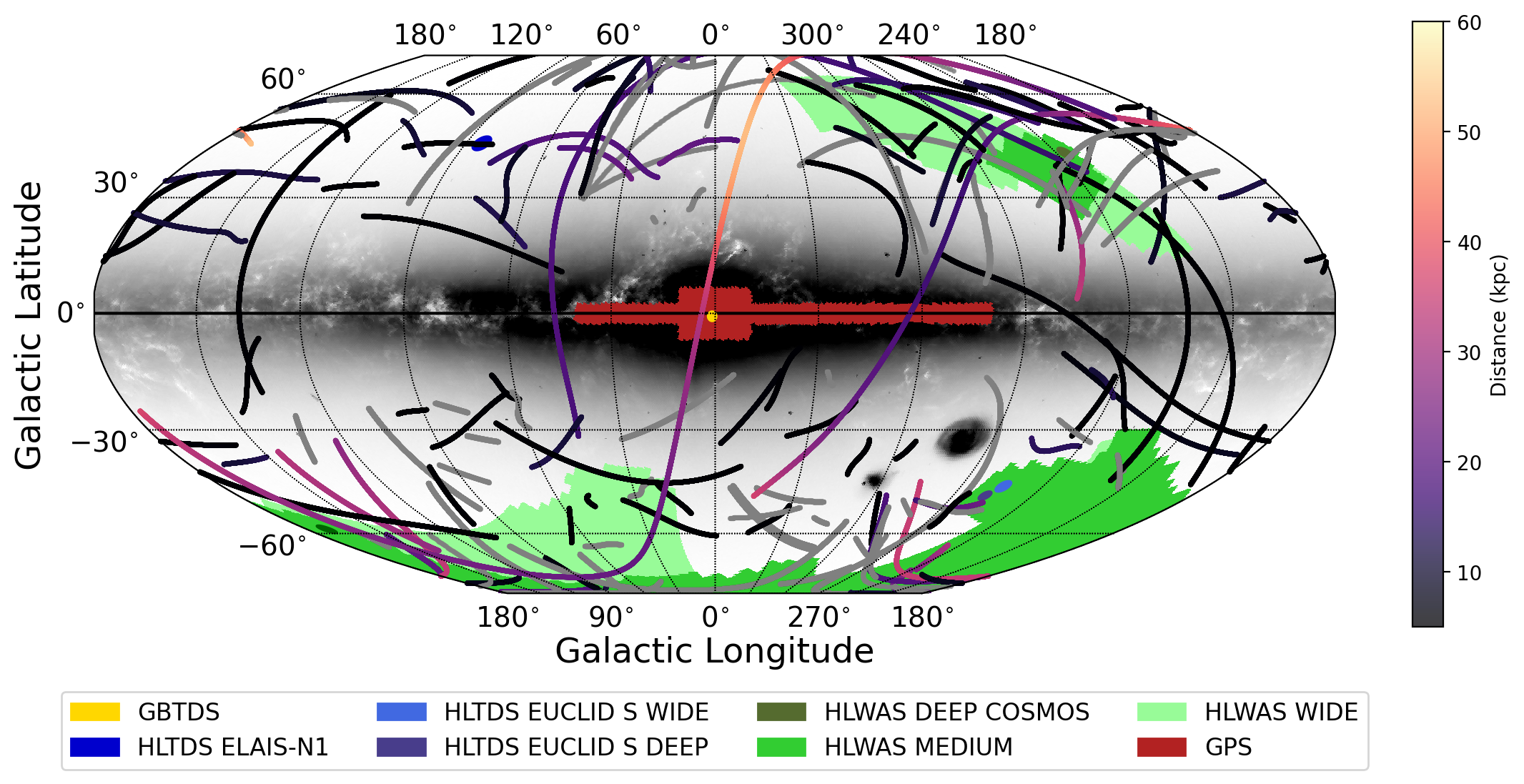}
\caption{Roman core community survey footprints and streams (colored by distance) shown in Galactic Coordinates on a Gaia map of the Milky Way. Light-green shows the single-band (F158) wide-tier HLWAS survey.  Darker green shows the 3-band (F106, F129, F158) medium-tier HLWAS.  Stream data from \texttt{galstreams} \citep{Mateu2023}.}
\label{fig:streams-allsky}
\end{center}
\end{figure*}

Fifty-nine of the known stellar streams pass through the footprint of the HLWAS, including 17 with no public proper motion information available and 23 with no distance measurements available in the \texttt{galstreams} database \citep{Mateu2023}, which is the main repository of information about Milky Way stellar streams. Dynamical modeling of streams, which requires this information, is the main route to constraining the overall mass distribution of the Milky Way outside the Galactic disk \citep[][and references therein]{2025NewAR.10001713B}. Most other characteristics used for comparisons to test dark matter (DM) theories, such as the halo's shape, density profile, ratio with the stellar mass of the MW, number and central densities of satellites, and population of dark substructures, are scaled by the MW mass. To date only a handful of known streams have been used to jointly constrain the MW's mass \citep[e.g.][]{Bovy_2016,Bonaca_2018,2021MNRAS.502.4170R,2023ApJ...945L..32C,Ibata2024}, mostly due to the lack of complete dynamical information---the primary source of accurate distances and proper motion measurements for streams is the Gaia mission \citep{2016A&A...595A...1G}, which is limited to magnitude 21 and brighter in the broad visible $G$ band used for astrometric measurements. At this magnitude, Gaia can measure proper motions accurately for individual giant stars to $\sim 50$ kpc from the Galactic center, and for more numerous main-sequence turnoff (MSTO) stars to $\sim 20$ kpc from the Galactic center. This is perhaps the most challenging regime in which to model the Milky Way's mass distribution, since the Large and Small Milky Clouds (at $\sim 45$ kpc galactocentric) must be modeled as separate mass distributions to obtain sensible answers \citep[e.g.][]{LawMajewski2010,2021MNRAS.501.2279V,2022arXiv220913663W,2023MNRAS.518..774L}. Avoiding the regime influenced by the MCs to obtain an accurate estimate for the MW's virial mass will require using streams with orbits outside $\sim 80-100$ kpc \citep{2021ApJ...919..109G,2023MNRAS.518..774L,2024ApJ...975..100G,2024ApJ...974..286A}. Currently the only stream with sufficient data at these distances is the Sagittarius stream \citep{2017ApJ...850...96H, 2021MNRAS.501.2279V}; the distances are mainly from standard candles (RR Lyr variable stars) which are abundant in this massive stream.

By contrast, the single-epoch depth of the HLWAS (Table \ref{tbl:survey}) will detect even individual MSTO stars beyond the edge of the MW halo. This could potentially make \emph{every} stream intersecting the HLWAS field available for dynamical modeling, an increase of more than \emph{ten times} the available sample, if we can realize the astrometric promise of the HLWAS for the near field.

\begin{figure*}[htbp]
\begin{center}
\includegraphics[height=0.45\textwidth]{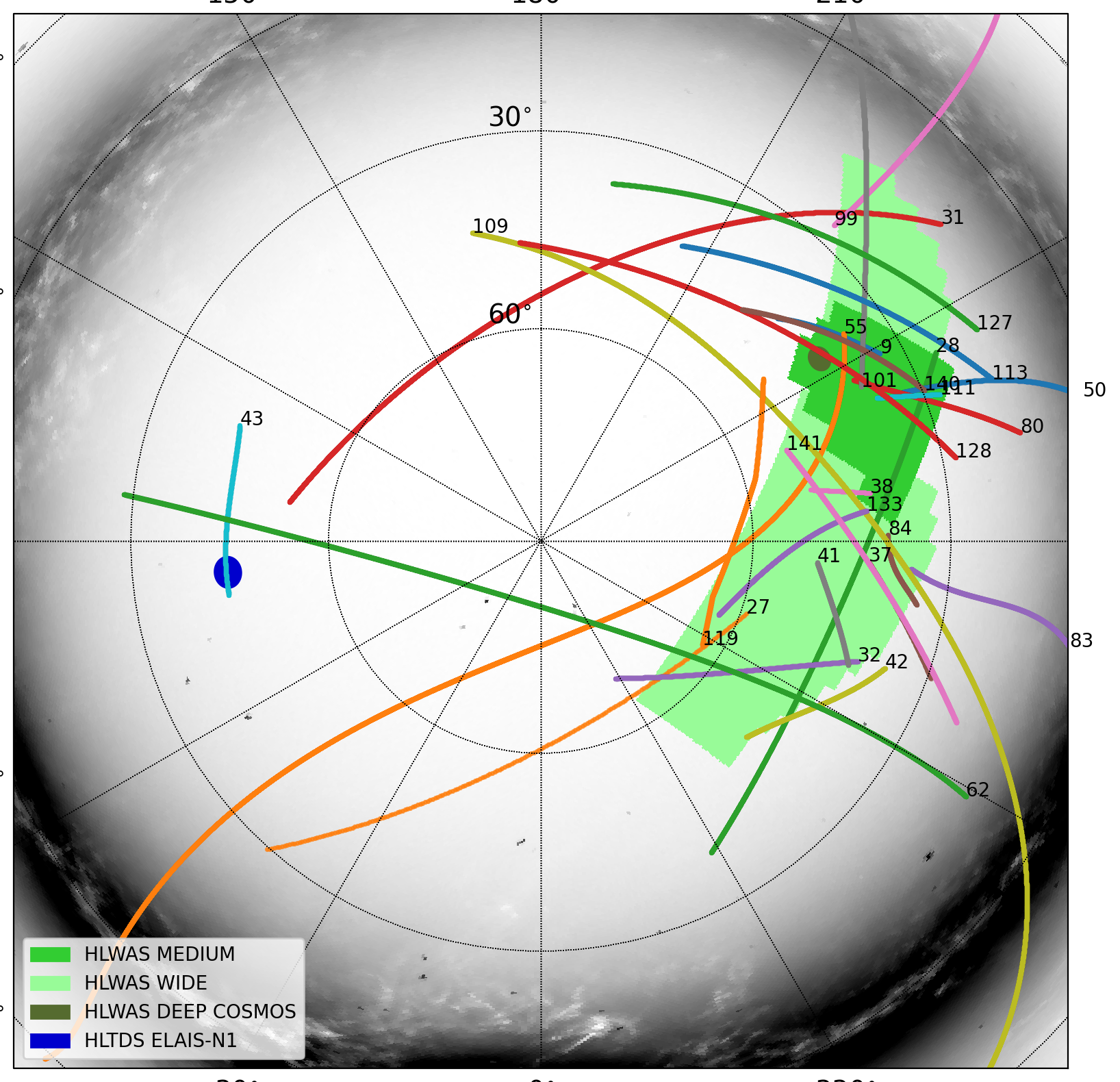} \includegraphics[height=0.45\textwidth]{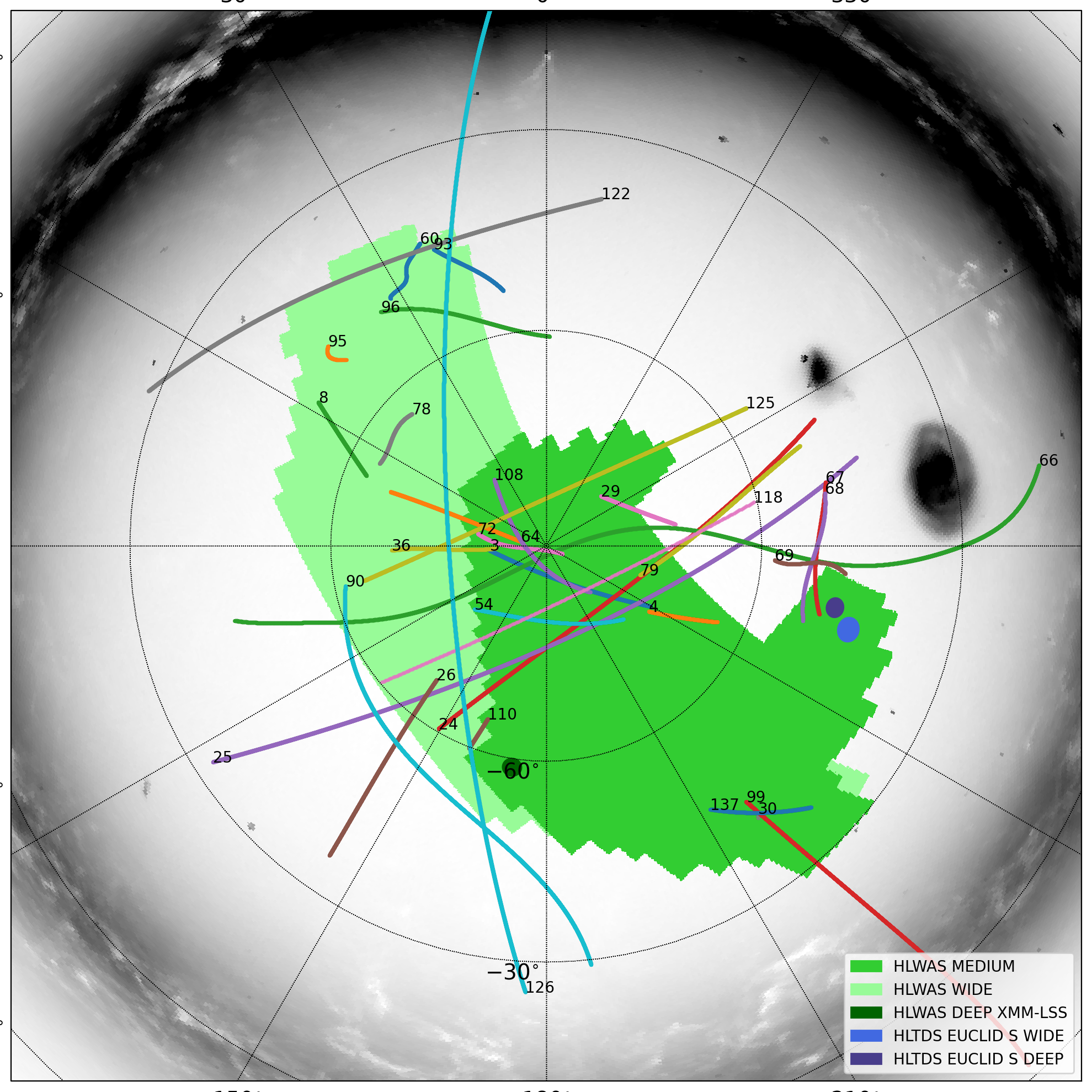}
\caption{Roman core community survey footprints and streams, north (left) and south (right) galactic cap views. Colors are the same as Figure~\ref{fig:streams-allsky}.  Stream data from \texttt{galstreams} \citep{Mateu2023}. Stream numbers correspond with Tables \ref{tbl:streams_north} (left) and \ref{tbl:streams_south} (right).}
\label{fig:streams-poles}
\end{center}
\end{figure*}

\begin{figure*}
    \centering
    \includegraphics[width=\linewidth]{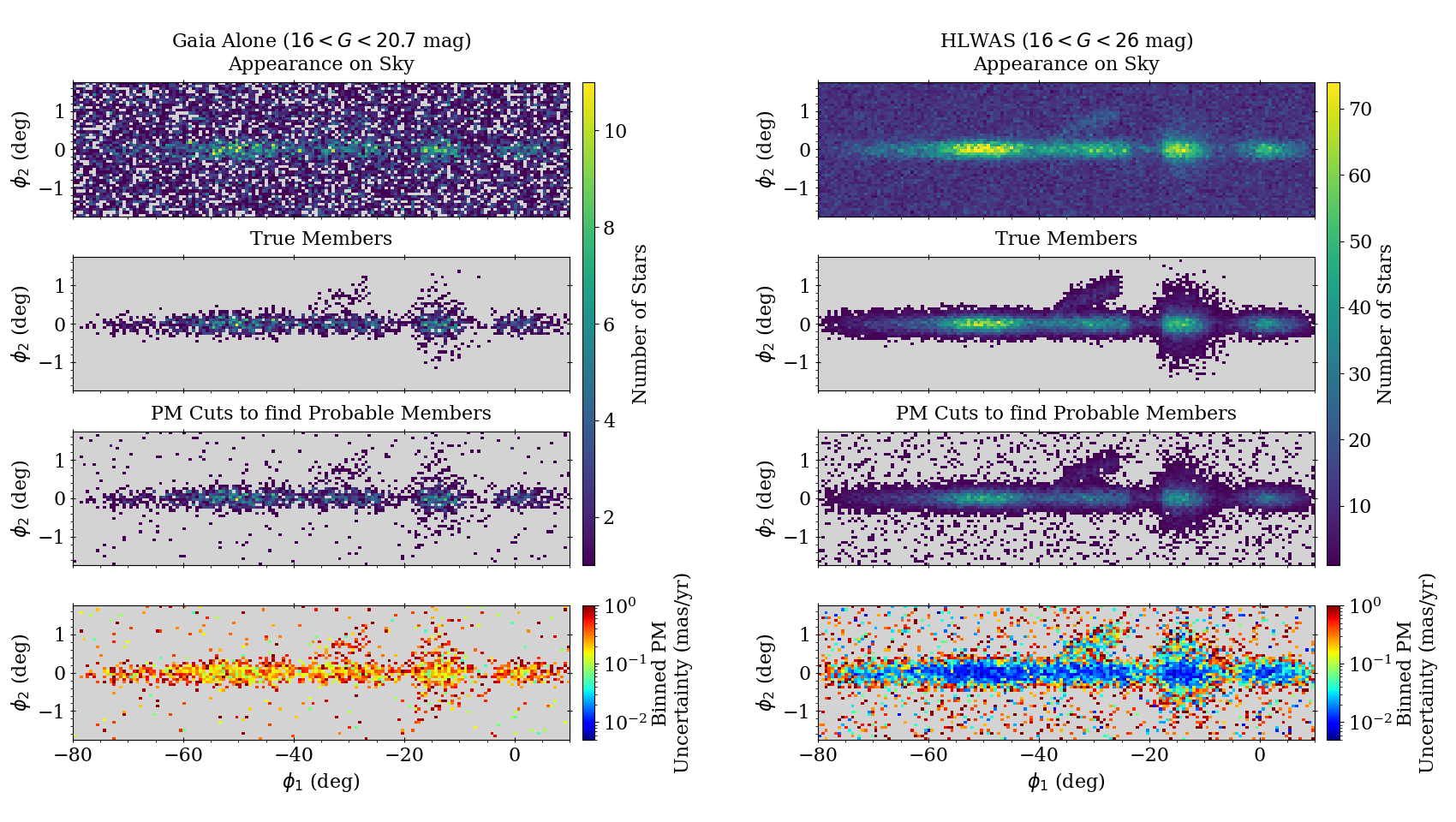}
    \caption{Realistic simulation of a stream against background MW halo and disk stars, with the axes shown in on- and off-axis stream coordinates. Stream properties -- such as heliocentric distance, width, age, [Fe/H], velocity dispersion, density fluctuations, and off-stream features -- are similar to those of GD-1. The left panels show the stream as viewed by Gaia DR3 while the right panels show the same region of sky as viewed by HLWAS-like observations. The top panels show a histogram of the stars (including MW stellar halo, the MW disk, and true stream members) in a common region of the sky when restricted to different magnitude limits. The second-from-top panels show all true member stars, while the third-from-top panels show the stars that survive a PM selection cut to identify likely members. By extending to much fainter magnitudes, Roman will be able to reveal hard-to-detect low-surface-brightness features, such as spurs ($\phi_1 \sim -30^\circ$, $\phi_2 \sim +1^\circ$), and kinematically heated regions ($\phi_1 \sim -15^\circ$, $-1 ^\circ <\phi_2 < +1^\circ$) in high resolution. The bottom panels show the binned PM uncertainty: by increasing the number of stars per bin and improving the individual PM precisions for faint stars, the HLWAS (using a 5 year baseline) could provide 10-30 times stronger kinematic constraints per area compared to Gaia alone.}
    \label{fig:gd1_mock}
\end{figure*}

Additionally, the increased depth in the HLWAS fields will not only greatly improve our ability to find new streams, but also to detect and characterize local density perturbations along known streams, which are one of only a few handles on the low-mass end of the present-day subhalo mass function and hence on the nature of DM \citep[][and references therein]{2022NatAs...6..897S,2022JHEAp..35..112B,2025NewAR.10001713B}. Roman's power in this regard is twofold: first, far more of the known streams will be observed down to the main-sequence turnoff (MSTO) where the density of stars increases sharply. Second, as we have already discovered with Gaia, improved proper motions down to the MSTO will allow for much cleaner removal of interlopers in the foreground and background of the stream. Combined, these effects make Roman a premier instrument for high-resolution stream mapping. Assuming proper motion precisions from HLWAS-like observations with a year 5 followup epoch (i.e. orange line of Figure~\ref{fig:Roman_astrometry_errs}), we compare the abilities of Gaia DR3 and Roman to detect a realistic stream in Figure~\ref{fig:gd1_mock}. By improving the per-star proper motion uncertainties as well as pushing to much fainter magnitudes, the HLWAS will be able to reveal previously-obscured substructure in high resolution and can potentially provide order-of-magnitude stronger kinematic constraints.

Several streams that fall in the HLWAS footprint are worth special mention: some, like Sagittarius (126, south), GD-1 (31, north), Orphan-Chenab (109, north), and Atlas-Aliqua Uma (3/4, south) are already known to be particularly scientifically interesting and have already been well-studied. Others are less well-studied but will get especially good coverage with Roman---their tracks lie entirely within the survey region, so the prospect of finding extensions for example is very high for these streams. Below we elaborate on a few potentially interesting cases.

\subsection{Sagittarius}

\begin{figure}
    \centering
    \includegraphics[width=\linewidth]{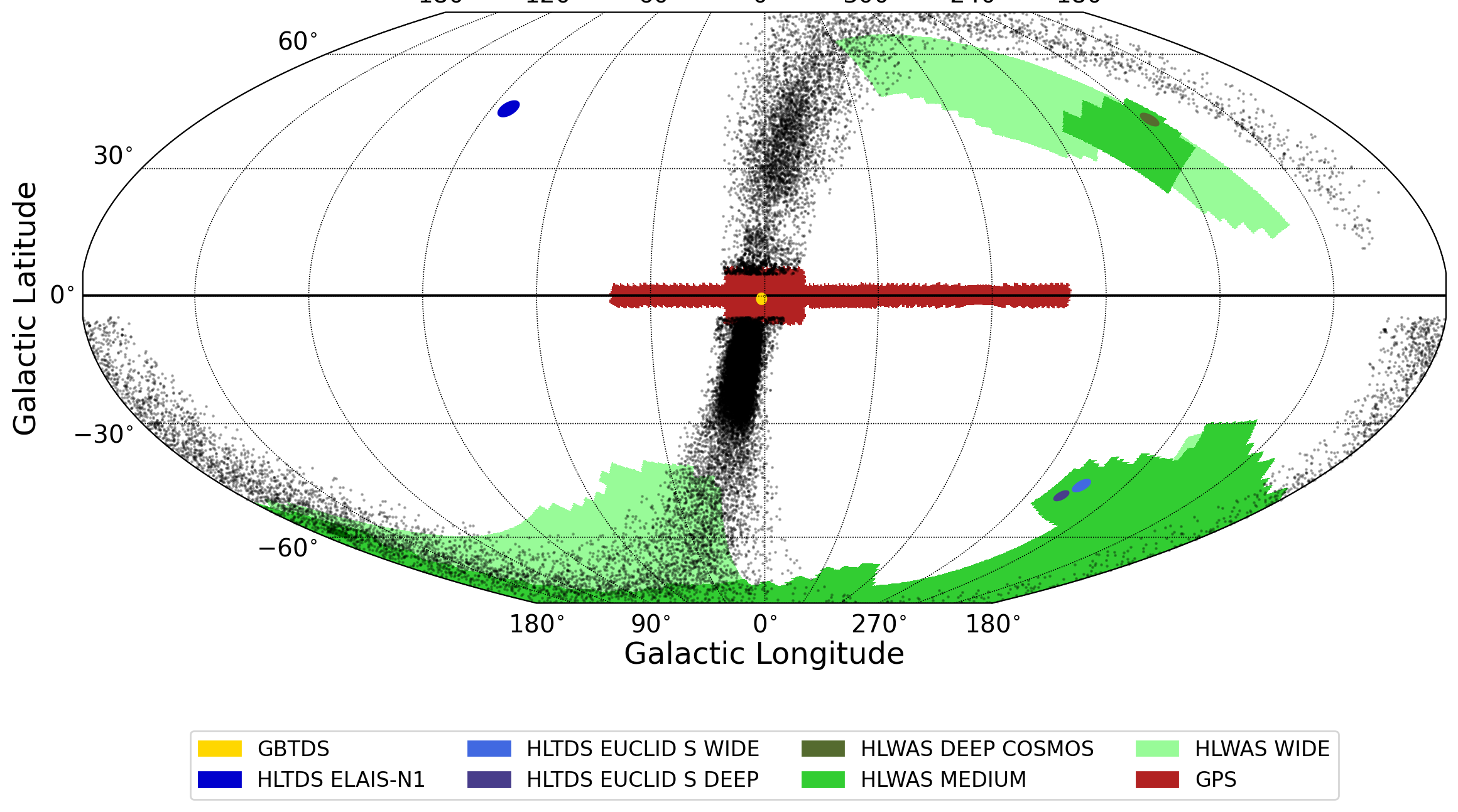}
    \caption{{\bf The Sagittarius stream wraps the Milky Way.}  Same as Figure~\ref{fig:streams-allsky}, but showing only the Sagitarius stream.}
    \label{fig:sgr-fullsky}
\end{figure}

\begin{figure*}
    \centering
    \begin{tabular}{cc}
         \includegraphics[width=0.5\linewidth]{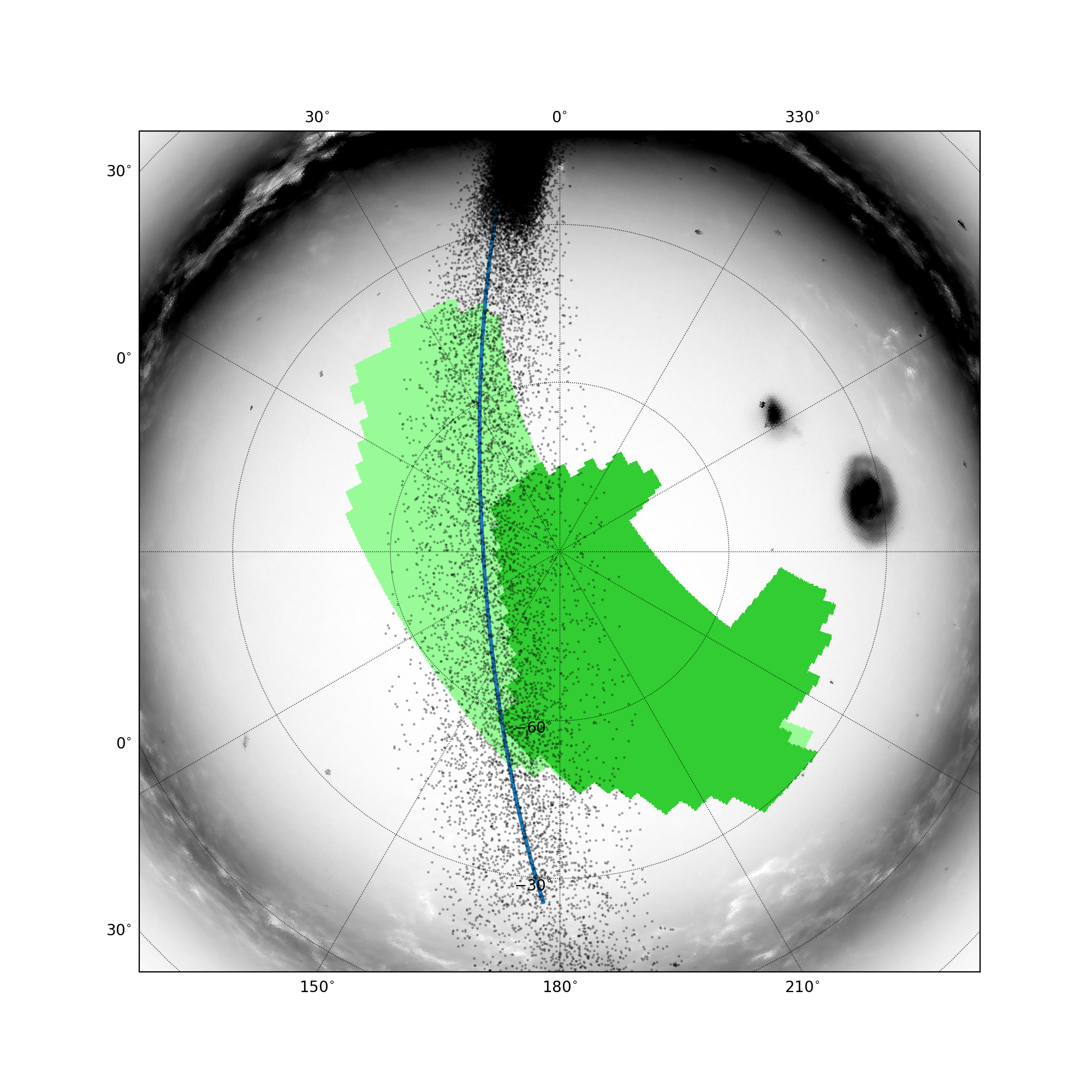}
    & 
         \includegraphics[width=0.5\linewidth]{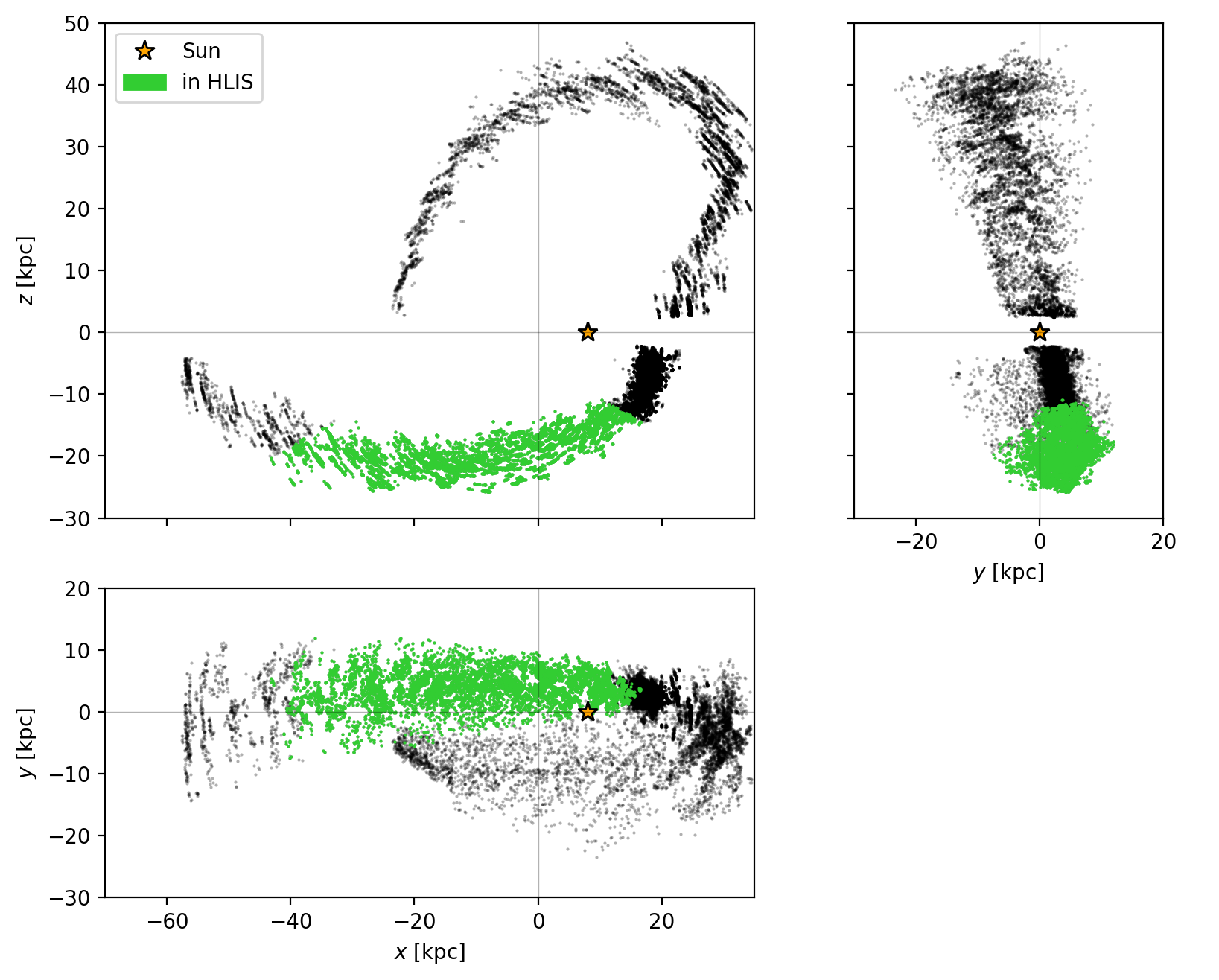}
    \end{tabular}
    \caption{{\bf The Sagittarius stream crosses the HLWAS footprint at apocenter.} \emph{Left:} view of south Galactic polar cap with Sgr stars (black points) crossing the HLWAS footprint (green). \emph{Right:} Three-dimensional Galactocentric view of the Sgr stream (black points) shows that the region viewed by the HLWAS (green) includes stars at one of the apocenters of the stream. The Sun's location in this projection is shown as an orange star.}
    \label{fig:sgr-south}
\end{figure*}

The longest and most massive stream in the Milky Way, the Sagittarius (Sgr) stream, wraps entirely around the Milky Way and passes through both the southern portion of the HLWAS and directly through the Galactic Plane Survey footprint (Figure \ref{fig:sgr-fullsky}). In the southern Galactic hemisphere, the HLWAS will view a significant portion of the stream near apocenter (Figure \ref{fig:sgr-south}). Proper motions at apocenter are particularly informative for measuring the degree of flattening or triaxiality of the MW's DM halo and the alignment of its approximate symmetry axes with the disk  \citep{Johnston2005,sanderson2013,Vera-Ciro_2013,belokurov2014sgr,2015MNRAS.454.2472H,reino2021,dongpaez2021}. Both these quantities are sensitive tests of DM theory, especially in the range of distances probed by this stream: in this radial range, the dominant symmetry of the global mass distribution is transitioning from being governed by the Galactic disk (at small Galactocentric radii) to the accreting filaments of the cosmic web (at large Galactocentric radii). This region is also where the wake left in the DM distribution by the MCs is expected to be strongest \citep{2020MNRAS.494L..11P,2021ApJ...919..109G, 2024ApJ...974..286A}. Since these transitions depend sensitively on the response of DM to dynamical perturbations, they make excellent tests of a nonzero DM cross-section.

\subsection{GD-1}

\begin{figure}
    \centering
    \includegraphics[width=0.5\textwidth]{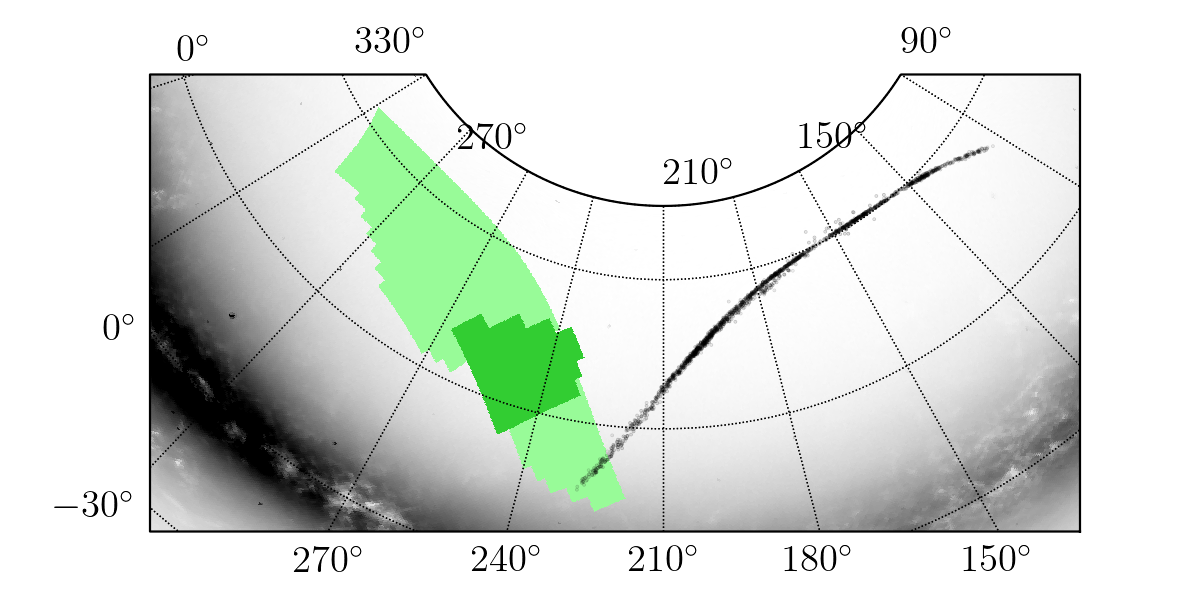}
    \caption{{\bf The GD-1 stream crosses the HLWAS Wide field at its eastern edge.} In this view in ICRS coordinates, the black points show probable GD-1 members (stars with $>50$\% membership probability and which pass a color-magnitude cut) from \citet{Tavangar2025}, colored by membership probability. The light and dark green shaded regions denote the wide and medium survey footprints respectively, as in previous figures. The background greyscale shows the Gaia DR3 flux map.}
    \label{fig:gd1-closeup}
\end{figure}

Thin streams in the HLWAS also offer the possibility of constraining the low-mass end of the substructure mass function, a highly diagnostic test of DM theories, through disturbances from putative interactions with these subhalos. These interactions affect the linear density of streams, creating unexpected gaps and off-track features, as well as their velocity dispersions by delivering small kicks to individual stars \citep[e.g.,][]{Johnston2002,Ibata2002,Yoon2011,Delos2022,Carlberg2024,Nibauer2025}. The most well-studied stream in this regard is GD-1 \citep{Grillmair2006_gd1,PriceWhelanBonaca2018_gd1,deBoer2018}, a thin, dynamically cold, high surface brightness stream in the northern sky. The eastern edge of GD-1 crosses the HLWAS above the Galactic plane (Table \ref{tbl:streams_north}) for $\approx 10$~deg, as seen in Figure~\ref{fig:gd1-closeup}. 

In this region, the HLWAS will significantly improve our measurement of the stream density and its velocity dispersion, two of the most critical properties for constraining stream interactions with subhalos \citep[e.g.]{bonaca19,Carlberg2024,Zhang_2025}. Increasing the depth from Gaia DR3's limit of $G\approx 21$~mag by $\approx 5$~mag will increase the number of observable GD-1 member stars in this region by a factor of $\approx 4$, assuming a Kroupa initial mass function \citep{Kroupa2001}. In absolute terms, \citet{Tavangar2025} found 66 probable\footnote{We define probable members as stars with a higher than 50\% membership probability according to \citet{Tavangar2025} that also pass their color-magnitude diagram cut.} GD-1 members in this region, meaning the HLWAS would contain an additional $\approx 200$. 
With only one band we would not be able to make color-magnitude cuts on these stars, so the HLWAS alone would likely not allow us to verify members fainter than the proper motion magnitude limit ($G\approx 24$~mag for Gaia-DR3 quality proper motions; see Section~\ref{sec:roman_PMs}). However, any additional observations, by either Roman or ground-based surveys like LSST, would permit adding color-magnitude selection to the classification of streams members. Even without this additional data, the HLWAS data will increase the number of GD-1 stars in the region covered by a factor of $\approx 2.5$.

The measured velocity dispersion of this part of GD-1 is $\approx 10$~mas/yr in each proper motion dimension, approximately 3 times as large as in its central section \citep{Tavangar2025}. However, due to the significantly lower number of identified member stars, this velocity dispersion measurement is not nearly as precise as it is for the center of GD-1. If the additional GD-1 members found with the HLWAS confirm the current proper motion dispersion measurements, it would imply that the outer edge of the stream has undergone significant heating over its orbital history. This could be the result of a few different phenomena, such as mass segregation or black-hole migration in the parent cluster \citep[e.g.,][]{Weatherford2026}, but one significant possibility is repeated subhalo interactions, through which we could place stronger constraints on the nature of DM \citep{Nibauer2026,Nibauer2025}. 

Lastly, some current estimates of GD-1's spatial extent suggest the stream ends within the HLWAS region (see Figure~\ref{fig:gd1-closeup}).
Therefore, while the HLWAS may not significantly extend GD-1, it could do so by a few degrees, allowing improved orbital fitting.

\subsection{Orphan-Chenab}

\begin{figure}
    \centering
    \includegraphics[width=0.5\textwidth]{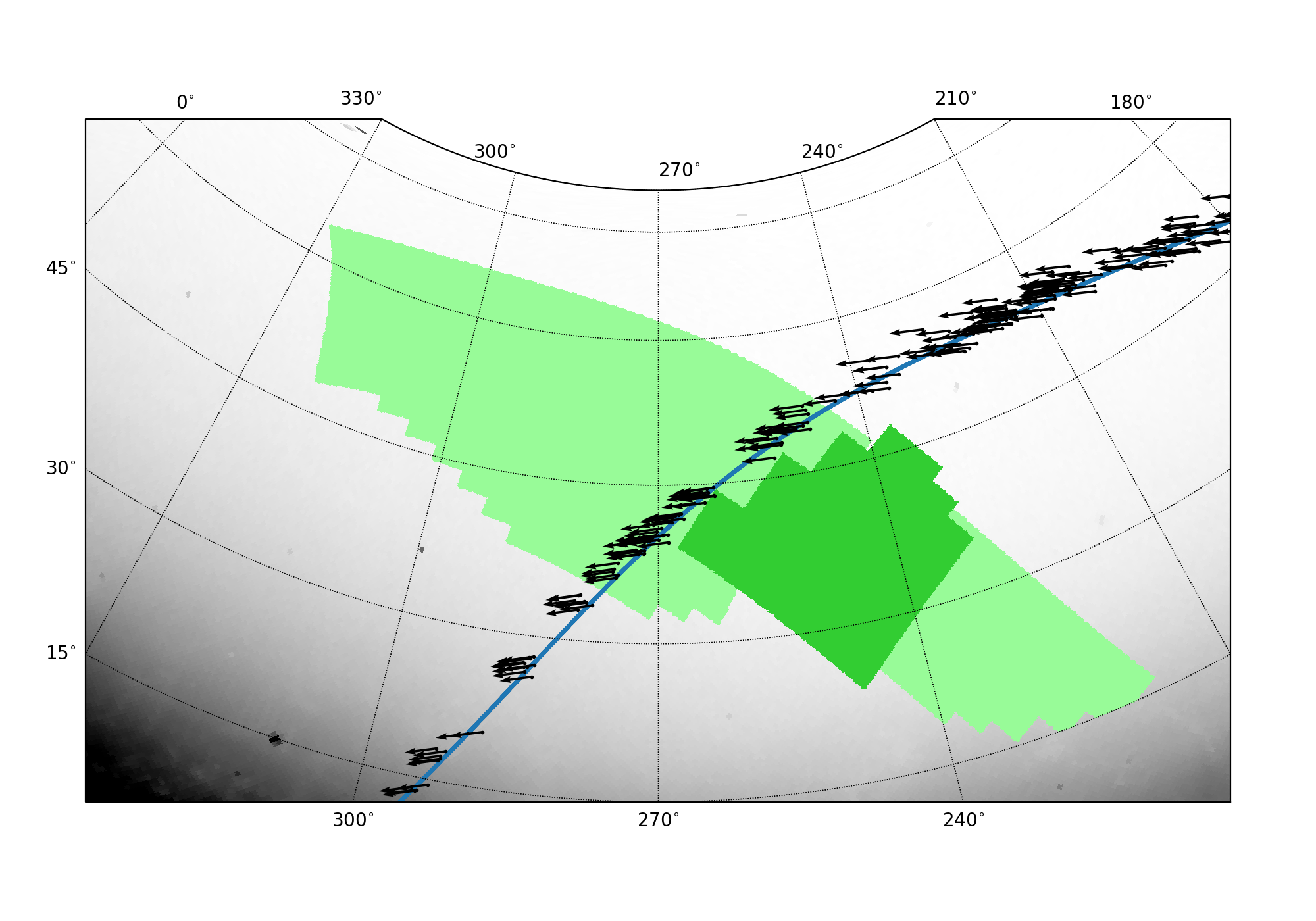}
    \caption{{\bf The Orphan-Chenab stream crosses the HLWAS field where the stream is being torqued by the Milky Clouds (MCs).} In this view in Galactic coordinates, black points with arrows show spectroscopic member stars in the stream from \citet{Koposov2023}; arrows denote direction of proper motion for each star (from Gaia) and the blue line denotes the stream track \citep{Koposov2019}. Torque from interaction with the MCs causes the proper motions to deviate from the direction of the track in the region of the HLWAS. Light and dark green regions denote the wide and medium survey footprints respectively. Background shading from  Gaia flux map.}
    \label{fig:oc-closeup}
\end{figure}

The Orphan-Chenab (OC) stream, a long thin stream from a disrupted dwarf galaxy, is perhaps our best probe of the region of the MW that is being deformed by the MCs. A portion of the stream displays incontrovertible evidence of the torque exerted by this ongoing merger: the velocities of the stream stars are nearly perpendicular to the track of the stream, instead of parallel as expected in a non-interacting system \citep{Koposov2019,Erkal2019,Koposov2023,2022arXiv220913663W,Brooks_2025}. The portion of the OC stream that crosses the HLWAS field contains the transition between the torqued and untorqued parts of the stream (Figure \ref{fig:oc-closeup}). The deeper photometry and proper motions that Roman can provide for this crucial region offer a unique opportunity to go beyond merely modeling the interaction and probe the wake of the MCs in the MW's DM halo \citep{2023MNRAS.518..774L}, a potentially powerful test of DM theories \citep{2023PhRvD.107d3014S}.

\subsection{Streams in the Galactic north}

In the north, 27 streams cross the HLWAS footprint (Table \ref{tbl:streams_north}; Figure \ref{fig:streams-poles} left), including a few of special interest. 
\begin{itemize}
    \item The {\bf Scamander} stream \citep[128,][]{Grillmair2017} crosses the northern deep field, and is thus likely to be observed very early in the survey. It will have extremely deep observations in mutiple filters by the end of the program. This stream currently does not have a publicly reported distance or proper motion, both of which should be possible to obtain with these observations.
    \item The {\bf Gaia-9} stream \citep[43,][]{Ibata2021} crosses the ELAIS survey field \citep{2000MNRAS.316..749O} which is part of the Roman High-Latitude Time Domain Survey. This field will have 5-band filter coverage at a range of cadences (see the ROTAC report for details), allowing, for example, searches for variable stars and high-precision proper motions (blue lines in Figure \ref{fig:HLWAS_Medium+HST_astrometry_errs}).
    \item The {\bf PS1} stream \citep[111/113,][]{Bernard2016} also has no reported distance or proper motion and will be in the HLWAS medium field, observed in 3 filters rather than 1 and thus likely earlier in the survey program.  
    \item The wide-field survey in the Galactic north encompasses the entire mapped length of four streams: {\bf Corvus} (28), {\bf Gaia-4} (38), {\bf Gaia-7} (41), and {\bf Sylgr} (133). The added depth of Roman will allow searches for extensions of these streams at either end, which if found can provide new constraints on the Milky Way's gravitational potential \citep[see, e.g.,][]{2015ApJ...799...28P,2023ApJ...954..215Y}.
\end{itemize}

\subsection{Streams in the Galactic south}

In the south, 32 streams cross the HLWAS footprint (Table \ref{tbl:streams_south}; Figure \ref{fig:streams-poles} right). Among these are five whose entire mapped length will be observed in three filters: {\bf ATLAS/Aliqa Uma} (AAU; 3/4), {\bf Orinoco} (108), \textbf{Turranburra/Eridanus} (137/30), \textbf{Kwando} (54), and \textbf{NGC288} (72). The AAU stream displays possible signatures of a past interaction with a substructure \citep[e.g.][]{2024arXiv240402953H,2025arXiv251207960N}. These streams are also excellent candidates for searches for extensions, as are several other streams in the wide-field footprint.

\begin{table*}[htp]
\caption{Streams crossing the HLWAS above the Galactic plane}
\begin{center}
\begin{tabular}{clccl}
\hline
\hline
ID & Name & distance? & PM? & Reference\\
\hline
9  &  C-10  &  *  & * & \citet{Ibata2024} \\
27  &  Cocytos  & & & \citet{Grillmair2009} \\
28  &  Corvus  & & & \citet{Mateu2018} \\
31  &  GD-1  &  *  & * & \citet{Ibata2021} \\
32  &  Gaia-1  &  *  & * & \citet{Ibata2021} \\
37  &  Gaia-3  & & & \citet{Malhan2018} \\
38  &  Gaia-4  & & & \citet{Malhan2018} \\
41  &  Gaia-7  &  *  & * & \citet{Ibata2021} \\
42  &  Gaia-8  &  *  & * & \citet{Ibata2021} \\
43  &  Gaia-9  &  *  & * & \citet{Ibata2021} \\
50  &  Jet  &  *  & * & \citet{Ferguson2022} \\
55  &  LMS-1  &  *  & * & \citet{Yuan2020} \\
62  &  M68-Fjorm  &  *  & * & \citet{Palau2019} \\
80  &  New-10  &  *  & * & \citet{Ibata2024} \\
83  &  New-13  &    & *  & \citet{Ibata2024} \\
84  &  New-14  &  *  & * & \citet{Ibata2024} \\
99  &  New-3  &  *  & * & \citet{Ibata2024} \\
101  &  New-5  &  *  & * & \citet{Ibata2024} \\
109  &  Orphan-Chenab  &  *  & * & \citet{Koposov2023} \\
111  &  PS1-B  & & & \citet{Bernard2016} \\
113  &  PS1-D  & & & \citet{Bernard2016} \\
119  &  Parallel  &  *  &  &\citet{Weiss2018} \\
127  &  Sangarius  & & & \citet{Grillmair2017} \\
128  &  Scamander  & & & \citet{Grillmair2017} \\
133  &  Sylgr  &  *  & * & \citet{Ibata2021} \\
140  &  Yangtze  &    & *  & \citet{Yang2023a} \\
141  &  Ylgr  &  *  & * & \citet{Ibata2021} \\
\hline
\hline

\end{tabular}
\end{center}
\tablecomments{In the distance and PM (proper motion) columns, a `*' indicates that this information is available in \texttt{galstreams}.}
\label{tbl:streams_north}
\end{table*}%

\begin{table*}[htp]
\caption{Streams crossing the HLWAS below the Galactic plane}
\begin{center}
\begin{tabular}{clccl}
\hline
\hline
ID & Name & distance? & PM? & Reference \\
\hline
3  &  AAU-ATLAS  &  *  & * & \citet{Li2021} \\
4  &  AAU-AliqaUma  &  *  & * & \citet{Li2021} \\
8  &  Aquarius  &  *  & * & \citet{Williams2011} \\
24  &  Cetus-New  &  *  & * & \citet{Yuan2021} \\
25  &  Cetus-Palca  &  *  & * & \citet{Thomas2021} \\
26  &  Cetus  &  *  &  &\citet{Yam2013} \\
29  &  Elqui  &    & *  & \citet{Shipp2019,Shipp2018} \\
30  &  Eridanus  & & & \citet{Myeong2017,Harris1996} \\
36  &  Gaia-2  &  *  & * & \citet{Ibata2021} \\
54  &  Kwando  &  *  & * & \citet{Ibata2021} \\
60  &  M30  & & & \citet{Sollima2020,Harris1996} \\
64  &  Molonglo  & & & \citet{Grillmair2017_south} \\
66  &  Murrumbidgee  & & & \citet{Grillmair2017_south} \\
67  &  NGC1261  &  *  & * & \citet{Ibata2021} \\
68  &  NGC1261a  &    & *  & \citet{Ibata2024} \\
69  &  NGC1261b  &  *  & * & \citet{Ibata2024} \\
72  &  NGC288  &  *  & * & \citet{Ibata2021} \\
78  &  NGC7492  &    & *  & \citet{Ibata2024} \\
79  &  New-1  &    & *  & \citet{Ibata2024} \\
90  &  New-2  &  *  & * & \citet{Ibata2024} \\
93  &  New-22  &  *  & * & \citet{Ibata2024} \\
95  &  New-24  &  *  & * & \citet{Ibata2024} \\
96  &  New-25  &  *  & * & \citet{Ibata2024} \\
99  &  New-3  &  *  & * & \citet{Ibata2024} \\
108  &  Orinoco  & & & \citet{Grillmair2017_south} \\
110  &  PS1-A  & & & \citet{Bernard2016} \\
118  &  Palca  & & & \citet{Shipp2018} \\
122  &  Phlegethon  &  *  & * & \citet{Ibata2021} \\
125  &  SGP-S  &    & *  & \citet{Yang2022} \\
126  &  Sagittarius  &  *  & * & \citet{Antoja2020,Ramos2020} \\
137  &  Turranburra  &    & *  & \citet{Shipp2019,Shipp2018} \\
\hline
\hline
\end{tabular}
\tablecomments{In the distance and PM (proper motion) columns, a `*' indicates that this information is available in \texttt{galstreams}.}
\end{center}
\label{tbl:streams_south}
\end{table*}%

\begin{figure*}[htbp]
\begin{center}
\includegraphics[width=\textwidth]{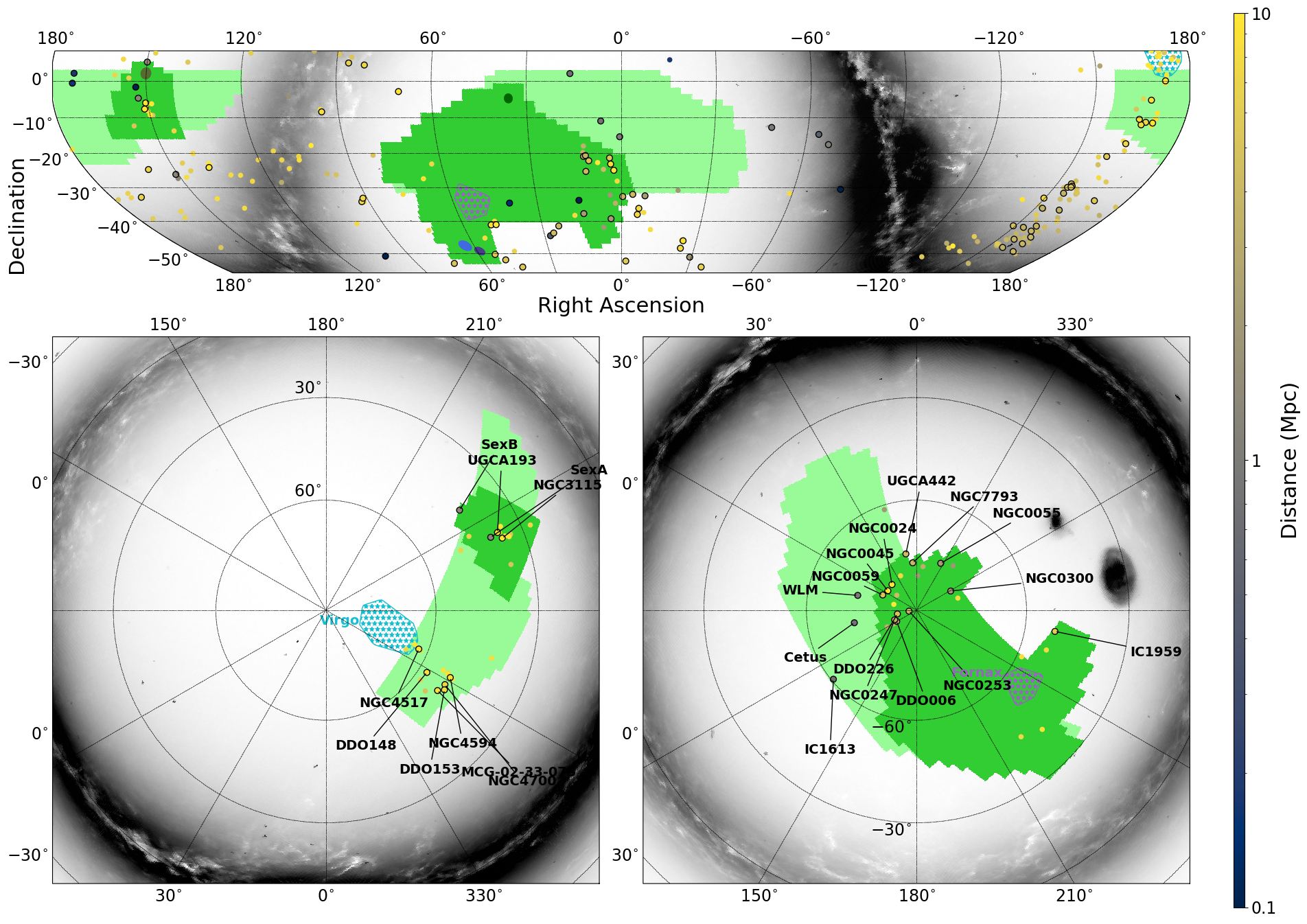}
\caption{{\it Top:} Roman core community survey footprints (HLWAS Medium in dark green, Wide in light green) and nearby galaxies (to 10 Mpc, colored by distance) shown in Equatorial coordinates (RA, Dec). Blue shades are MW satellites, gray are Local Group, and yellow are Local Volume. Galaxies in black circles have major-axis diameter larger than 2 arcmin. The Virgo galaxy cluster is shown with a blue polygon and the Fornax galaxy cluster is shown with a gray polygon.  Galaxy data from \citet{karachentsev2013}.  {\it Bottom:} Same as top, but shown in Galactic polar coordinates to allow space to label galaxies and clusters. }
\label{fig:nbgs-allsky}
\end{center}
\end{figure*}

\section{Nearby Galaxies}
\label{sec:nbgs}

A total of 77 galaxies within 10 Mpc are in the HLWAS footprint, including fourteen Milky Way satellites, five members of the Local Group (LG), and the Sculptor Group of galaxies (Tables \ref{tbl:nbgs1}--\ref{tbl:nbgs3}, Figure \ref{fig:nbgs-allsky}). Of these, 10 satellites/LG members and 37 Local Volume galaxies will be in the HLWAS Medium field, which observes in three bands (Table \ref{tbl:survey}). This multicolor imaging will give us an unprecedented view of galaxy formation.

\subsection{Stellar Populations}

Detailed and significant constraints on the nature of DM as well as the formation and evolution of galaxies can be gleaned from studies of the resolved stars in the nearby galaxies within the HLWAS footprint.  The virial radii of these galaxies are large on the sky, and as a result their diffuse stellar halos have yet to be characterized.  Within the LSST footprint, the Sculptor Group coverage in particular can enable resolved stellar populations studies in addition to the planned cosmology measurements.  These observations will reach low surface brightness regions through resolved star counts.  In addition, in these regions we can leverage LSST full depth photometry in these regions for panchromatic spectral energy distribution fitting of the stars resolved by Roman.  

The $\Lambda$-cold dark matter ($\Lambda$CDM) cosmological model and its viable alternatives appear equally successful in reproducing the large-scale ($>$1 Mpc) distribution of galaxies \citep[e.g.][]{Planck20}. The main variance between viable DM models is on very small scales, in the number and properties of low-mass galaxy halos \citep{Bullock17,2022NatAs...6..897S}. Numerical simulations predict that `cold dark matter' (CDM) forms gravitationally bound structures (called halos) far below the mass needed to host a galaxy \citep[$M_{\mathrm{halo}} \sim 10^7M_{\odot}$;][]{2006MNRAS.371..395B,2009ApJ...693.1859B,Bovill11,Bullock17}, implying the existence of a multitude of dark substructures around galaxies. 

The entire HLWAS footprint, but especially the areas around nearby galaxies, will provide a new sensitive data set for finding and characterizing faint dwarf satellites.
Dwarf and satellite galaxies, in particular ultra-faint dwarf galaxies \citep[M$_V>$$-$7.7; ][]{Simon2019}, are the primary observational probe of low-mass DM halos \citep{Moore99}. The smallest of these halos (and their associated stars) are predicted to have formed very early \citep[e.g.][]{Bullock00,Bovill11, Wheeler15,Jeon17,Applebaum21}, making them incredibly sensitive to the joint physics of DM, galaxy formation and reionization.

Of all upcoming observatories, Roman promises the most progress in this critical area---even deep lensed JWST fields cannot resolve the precursors to ultra-faint dwarf galaxies \citep{Weisz2017}. 
Roman should recover and map the extended structures of ultra-faint dwarf galaxies out to a distance of $\sim$5 Mpc---a thousandfold increase in volume---both as massive galaxy satellites
and in the field \citep{Bell2021}, revolutionizing our ability to jointly constrain on galaxy formation physics, reionization and DM.

For galaxies within the HLWAS, we will be able to detect streams in their extended halos.
As the tidally stripped remnants of accreted galaxies, stellar streams and other substructure encode both the accretion history of the host galaxy and the evolution of low-mass galaxies \citep[][Starkenburg et al.\ in prep.]{1999Natur.402...53H, 1999AJ....118.1719J, 2001ApJ...557..137J}. Additionally, streams from dwarf galaxies or globular clusters (GCs) are dynamical tracers of the host halo mass, and 
thin stellar streams from GCs indirectly probe encounters with DM halos too small to host galaxies. To date, 
dynamically cold globular cluster streams have only been detected and mapped in the MW \citep[e.g.,][]{deBoer2018,PriceWhelan2018,Bonaca2020}. However, Roman will be sensitive to these streams out to $\sim$6 Mpc \citep{Pearson2019,Pearson2022a,Nibauer_2023, Walder_2025, 2025arXiv250802666N, 2026arXiv260115373C} and to stream gaps out to $\sim$2--3 Mpc \citep{2023arXiv230512045A}, allowing for the first time statistical comparisons of stream structure to predictions from various DM candidates \citep{2022NatAs...6..897S}. Roman will also be able to detect streams around lower-mass galaxies, for which the accretion rate is poorly constrained \citep{2026A&A...707L...1S}.

Furthermore, the resolved photometry allows reliable constraints on the galaxies' star formation histories. A galaxy’s star formation history (SFH) describes its rate of star formation and metal enrichment over cosmic time, telling the story of how the galaxy formed and evolved. SFHs reconstructed from resolved stellar populations are the current gold standard approach 
\citep[e.g.,][]{2005ARA&A..43..387G, Conroy2013, Annibali2022}. 
With proper planning, Roman's  
potential to cover hundreds of galaxies from the Local Group to $\sim$10 Mpc could yield SFHs 
over a broader range of galaxy types than has ever been attempted, with coverage that can measure population gradients and structural variations.  HST and JWST have delivered exquisite SFHs, but the restrictive FoV has limited this work almost exclusively to dwarf galaxies \citep[e.g.,][]{McQuinn2010, Weisz2011, Skillman2017, 2023ApJS..268...15W, 2024ApJ...961...16M,2024ApJ...976...60M}. In this limited regime, 
the small area probed may not be representative of the galaxy as a whole \citep{Graus2019}.

The stellar halo structure then also allows constraints on the merger history. 
In addition to the growth histories of the galaxies themselves, Roman will provide an unprecedentedly detailed view of their diffuse outskirts. As galaxies merge, much of the resulting debris is deposited on orbits that extend to large galactocentric distances, forming an extended, richly-structured stellar halo \citep{Bullock00,Bullock2005,Cooper2010}.  
The largest individual mergers deposit the most debris and therefore dominate measurements of the stellar mass, metallicity, and SFH of stellar halos \citep{Cooper2010,Deason2016,DSouza2018a}. 
Halo {\it metallicities} measured using the metal-sensitive colors of red giant branch stars \citep{Monachesi2016}, complemented where possible using ground-based spectroscopic metallicities for `bright' stars identified in Roman imaging \citep{Toloba2016}, can constrain the time evolution of the halo mass--metallicity relation, while halo {\it SFHs} can 
constrain the merger time \citep{DSouza2018b,Harmsen2021,Panithanpaisal2021}. The {\it structure} of stellar halos reflects primarily recent mergers whose debris has not yet had time to phase mix \citep{Johnston2008,Panithanpaisal2021}, probing the galaxy's recent interaction history and providing the stellar streams that can map its DM halo \citep[e.g.,][]{Fardal2013, Pearson2022}. 

The scientific promise of these techniques has only been partially realized for the MW and M31, revealing many streams \citep{belokurov2006,Malhan2018,Shipp2018} as well as massive mergers in the MW $\sim$4--9 Gyr ago \citep{Helmi2018,Belokurov2018,2022ApJ...932L..16D} and in M31 $\sim$2 Gyr ago \citep{2001Natur.412...49I,DSouza2018b} that likely had profound influence on the galaxies' disks \citep[e.g.][]{Dalcanton2015,Williams2017,Hammer2018,2019MNRAS.483.1427L,Belokurov2020,2021MNRAS.508.1459H,2022MNRAS.516L...7H}. 
The HLWAS, and hopefully a more dedicated nearby galaxies survey, will extend these insights to a statistically-meaningful sample of nearby galaxies, connecting galaxy properties/SFHs with satellite populations, DM halo masses and merger histories for the very first time, offering the first-ever comprehensive test of galaxy formation models.

Nearby galaxies in the HLWAS can also be used to test for systematics in the extragalactic distance scale, particularly effects of population gradients within a galaxy on its measured distance. Attempts at these tests with HST and JWST \citep[e.g.][]{2023ApJ...954L..31S,2025ApJ...985..182L} have had mixed success thanks to the limited areal coverage of the disk and halo possible with smaller fields of view. Roman will thus offer a new opportunity to solidify these important first rungs on our distance ladder.

\subsection{Milky Way Satellites}

\begin{figure*}
    \begin{center}
        \includegraphics[width=\textwidth]{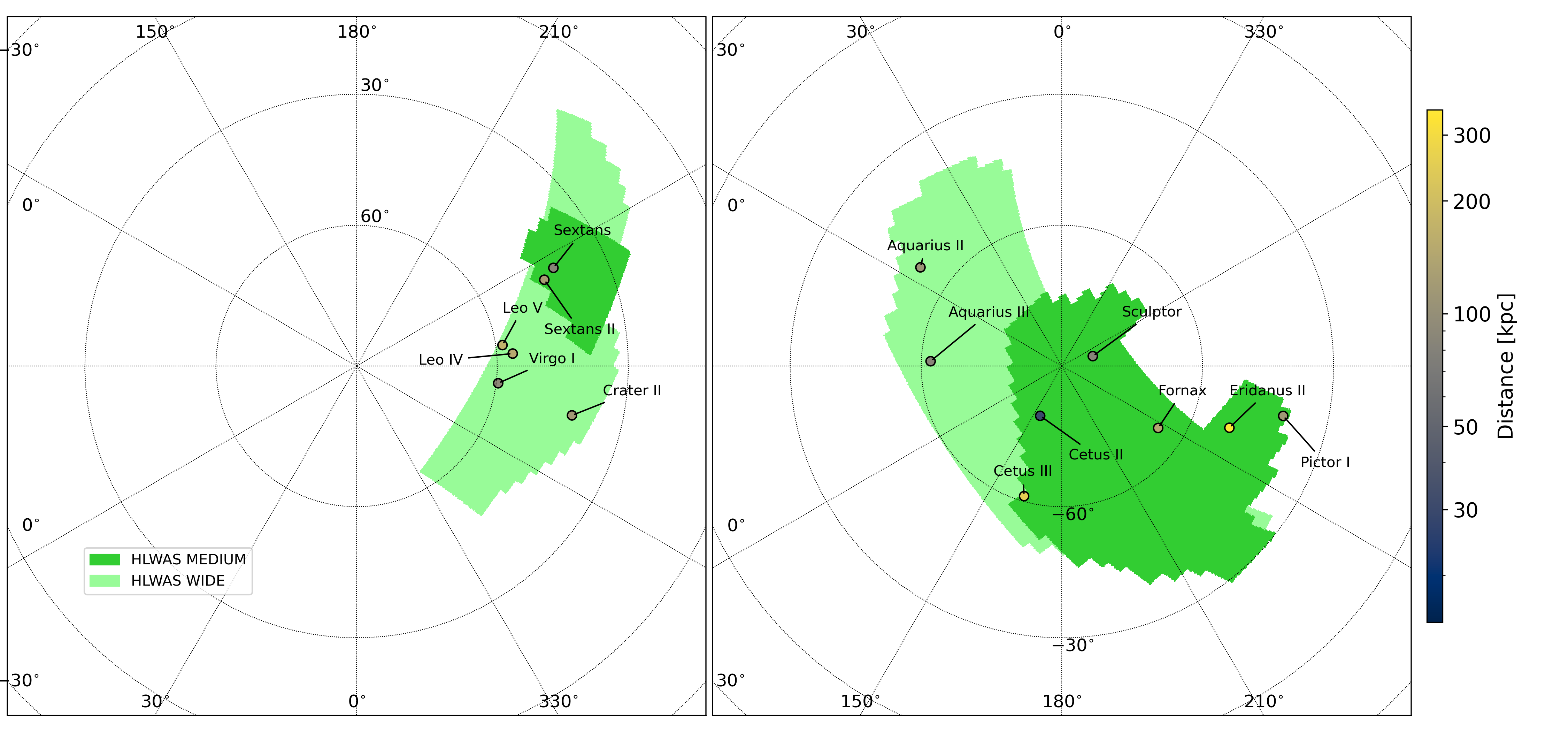}
    \end{center}
    \caption{Satellite galaxies of the Milky Way in the core community survey footprints, shown in galactic coordinates near the North Galactic Pole region (left) and South Galactic Pole region (right). Color scale indicates distance. Information for these galaxies is in Table \ref{tbl:nbgs1}. \label{fig:sats}}
\end{figure*}

Fourteen known Milky Way satellites are in the footprint of the HLWAS (Figure \ref{fig:sats}; Table \ref{tbl:nbgs1}). Many of these (including Sculptor, Fornax, Sextans, Leo IV, Leo V) have been observed by HST with time-baselines of up to 30 years, offering the prospect of proper motion uncertainties below 0.1 mas/yr per star (Figure \ref{fig:HLWAS_Medium+HST_astrometry_errs}) and hence access to the internal dynamics of these galaxies in full 3D \citep{2024ApJ...970....1V}. Three-dimensional analysis breaks the significant mass-anisotropy degeneracy in one-dimensional mass modeling of these galaxies, which has historically limited their constraining power for DM models \citep{2007ApJ...657L...1S}. The HLWAS point-source depth for the medium and wide surveys is comparable to the depth commonly attained by HST observations of many of these systems for modeling their star formation histories \citep[e.g.][]{2014ApJ...789..147W} but is significantly shallower than JWST has reached \citep[e.g.][]{2023ApJS..268...15W}. The clear advantage of Roman is its field of view: where Webb can offer deep views of a few representative fields, Roman is ideal for mapping the extended outskirts of these systems, for example to search for evidence of tidal features, to create spatially resolved SFHs across galaxies, and as a test of the degree to which the smaller fields are indeed representative. 

\begin{table*}[htp]
\caption{Nearby galaxies in the HLWAS fields: MW satellites and Local Group}
\begin{center}
\begin{tabular}{lcccccl}
\hline
\hline
Name & RA & Dec & $M$ & D [Mpc] & $m$ & which survey field\\
\hline
Cetus II & 01 17 52.8 & -17 25 12.0 & 0.0 & 0.030 & 17.4 & medium S \\
Sculptor & 01 00 04.4 & -33 43 07.0 & -10.8 & 0.084 & 8.8 & medium S \\
Aquarius III & 23 48 52.3 & -03 29 20.4 & -2.5 & 0.086 & 17.2 & wide S \\
Sextans & 10 13 04.3 & -01 37 12.0 & -8.7 & 0.086 & 10.9 & medium N \\
Virgo I & 12 00 09.1 & +00 40 51.6 & -0.9 & 0.091 & 18.9 & wide N \\
Aquarius II & 22 33 55.5 & -09 19 38.6 & -4.4 & 0.108 & 15.8 & wide S \\
Pictor I & 04 43 47.4 & -50 16 59.0 & -3.1 & 0.115 & 17.2 & medium S \\
Crater II & 11 49 14.4 & -18 24 46.8 & -8.2 & 0.117 & 12.2 & wide N \\
Sextans II & 10 25 44.9 & +00 37 51.6 & -3.9 & 0.126 & 16.6 & medium N \\
Fornax & 02 39 50.0 & -34 29 58.9 & -13.4 & 0.143 & 7.4 & medium S \\
Leo IV & 11 32 57.7 & +00 32 43.1 & -5.0 & 0.151 & 15.9 & wide N \\
Leo V & 11 31 08.6 & +02 13 09.8 & -4.4 & 0.169 & 16.7 & wide N \\
Cetus III & 02 05 19.4 & -04 16 12.0 & -3.5 & 0.251 & 18.6 & medium S \\
Eridanus II & 03 44 22.2 & -43 31 58.4 & -7.1 & 0.370 & 15.7 & medium S \\
\hline
IC1613 & 01 04 47.8 & +2 07 60.0 & -14.5 & 0.730 & 7.4 & wide S \\
Cetus & 00 26 11.0 & -11 02 40.0 & -10.2 & 0.780 & 10.2 & wide S \\
WLM & 00 01 58.1 & -15 27 40.0 & -14.1 & 0.970 & 9.0 & wide S \\
SexA & 10 11 0.8 & -4 41 34.0 & -13.9 & 1.320 & 10.2 & medium N \\
SexB & 10 00 0.1 & +5 19 56.0 & -14.0 & 1.360 & 9.5 & medium N \\

\hline
\hline
\end{tabular}
\end{center}
\tablecomments{Milky Way satellite data from \citet{2025OJAp....8E.142P} and references therein, courtesy of E. Vitral and A. Pace; other data from \citet{karachentsev2013}. Consequently, absolute magnitudes are $M_V$ for Milky Way satellites and $M_B$ for others; apparent mags are $m_V$ for satellites and $m_K$ for others. }
\label{tbl:nbgs1}
\end{table*}%

\begin{table*}[htp]
\caption{Nearby galaxies in the HLWAS fields: outside the Local Group to 8 Mpc.}
\begin{center}
\begin{tabular}{lcccccl}
\hline
\hline
Name & RA & Dec & $M_B$ & D [Mpc] & $m_K$ & which survey field\\
\hline
ESO410-005 & 00 15 31.4 & -32 10 48.0 & -11.6 & 1.92 & 12.5 & medium S \\
ESO294-010 & 00 26 33.3 & -41 51 20.0 & -10.9 & 1.92 & 14.1 & medium S \\
KK258 & 22 40 43.9 & -30 47 59.0 & -10.3 & 2.00 & 12.2 & wide S \\
NGC0055 & 00 15 8.5 & -39 13 13.0 & -18.4 & 2.13 & 6.2 & medium S \\
NGC0300 & 00 54 53.5 & -37 40 57.0 & -17.9 & 2.15 & 6.4 & medium S \\
ESO349-031 & 00 08 13.3 & -34 34 42.0 & -11.9 & 3.21 & 13.0 & medium S \\
DDO006 & 00 49 49.3 & -21 00 58.0 & -12.4 & 3.34 & 13.3 & medium S \\
KDG002 & 00 49 21.1 & -18 04 28.0 & -11.4 & 3.40 & 13.9 & medium S \\
ESO540-032 & 00 50 24.6 & -19 54 25.0 & -11.3 & 3.42 & 14.0 & medium S \\
NGC0247 & 00 47 8.3 & -20 45 36.0 & -18.5 & 3.65 & 7.4 & medium S \\
NGC7793 & 23 57 49.4 & -32 35 24.0 & -18.5 & 3.91 & 6.8 & medium S \\
NGC0253 & 00 47 34.3 & -25 17 32.0 & -21.3 & 3.94 & 3.8 & medium S \\
Sc22 & 00 23 51.7 & -24 42 18.0 & -10.5 & 4.21 & 13.6 & medium S \\
UGCA442 & 23 43 46.0 & -31 57 33.0 & -14.7 & 4.27 & 11.4 & medium S \\
DDO226 & 00 43 3.8 & -22 15 1.0 & -13.6 & 4.92 & 12.5 & medium S \\
KDG218 & 13 05 44.0 & -7 45 20.0 & -11.9 & 5.00 & 12.5 & wide N \\
KKSG19 & 10 24 28.3 & -12 25 57.0 & -12.1 & 5.11 & 14.2 & medium N \\
LV J0956-0929 & 09 56 37.6 & -9 29 11.0 & -13.0 & 5.18 & 13.3 & medium N \\
NGC0059 & 00 15 25.1 & -21 26 38.0 & -15.7 & 5.30 & 10.1 & medium S \\
IC1959 & 03 33 11.8 & -50 24 38.0 & -16.0 & 6.05 & 11.0 & medium S \\
HIPASS J1258-04 & 12 58 49.6 & -4 53 19.0 & -13.9 & 6.50 & 12.8 & wide N \\
LV J1021+0054 & 10 21 38.9 & +0 54 0.0 & -11.9 & 6.78 & 14.9 & medium N \\
UGC05797 & 10 39 25.2 & +1 43 7.0 & -15.1 & 7.00 & 11.8 & wide N \\
NGC4700 & 12 49 7.6 & -11 24 41.0 & -17.1 & 7.30 & 10.5 & wide N \\
LV J0935-1348 & 09 35 21.6 & -13 48 52.0 & -12.9 & 7.30 & 14.1 & medium N \\
NGC4600 & 12 40 23.0 & +3 07 4.0 & -15.8 & 7.35 & 9.8 & wide N \\
ESO483-013 & 04 12 41.1 & -23 09 32.0 & -15.5 & 7.40 & 10.8 & medium S \\
ESO409-015 & 00 05 31.8 & -28 05 53.0 & -14.4 & 7.70 & 12.7 & medium S \\
KKSG29 & 12 37 14.1 & -10 29 51.0 & -13.0 & 7.70 & 14.1 & wide N \\
PGC013294 & 03 35 56.8 & -45 11 29.0 & -13.4 & 7.98 & 13.7 & medium S \\
DDO153 & 12 53 57.5 & -12 06 31.0 & -15.2 & 8.00 & 12.0 & wide N \\
\hline
\hline
\end{tabular}
\tablecomments{Data from \citet{karachentsev2013}. }
\end{center}

\label{tbl:nbgs2}
\end{table*}

\begin{table*}[htp]
\caption{Nearby galaxies in the HLWAS fields: 8 to 10 Mpc.}
\begin{center}
\begin{tabular}{lcccccl}
\hline
\hline
Name & RA & Dec & $M_B$ & D [Mpc] & $m_K$ & which survey field\\
\hline
KDG155 & 12 33 8.0 & +0 31 59.0 & -13.2 & 8.11 & 14.0 & wide N \\
AM0106-382 & 01 08 22.0 & -38 12 33.0 & -13.4 & 8.2 & 13.9 & medium S \\
ESO572-034 & 11 58 58.1 & -19 01 48.0 & -15.6 & 8.5 & 11.5 & wide N \\
MCG-02-33-075 & 12 58 28.3 & -10 34 37.0 & -15.2 & 8.7 & 11.8 & wide N \\
ESO300-016 & 03 10 10.5 & -40 00 11.0 & -14.2 & 8.8 & 13.2 & medium S \\
DDO148 & 12 48 43.1 & -5 15 14.0 & -16.2 & 9.0 & 11.0 & wide N \\
KKSG30 & 12 37 35.9 & -8 52 2.0 & -13.6 & 9.1 & 13.8 & wide N \\
NGC1592 & 04 29 40.8 & -27 24 31.0 & -15.5 & 9.1 & 12.7 & medium S \\
NGC0045 & 00 14 3.9 & -23 10 56.0 & -18.5 & 9.2 & 9.1 & medium S \\
KKSG33 & 12 40 8.9 & -12 21 53.0 & -11.5 & 9.3 & 14.2 & wide N \\
KKSG34 & 12 41 18.9 & -11 55 39.0 & -12.1 & 9.3 & 13.7 & wide N \\
KKSG32 & 12 39 55.0 & -11 44 48.0 & -11.6 & 9.3 & 14.2 & wide N \\
SUCD1 & 12 40 3.1 & -11 40 4.0 & -11.5 & 9.3 & 14.7 & wide N \\
NGC4594 & 12 39 59.1 & -11 37 23.0 & -21.8 & 9.3 & 5.0 & wide N \\
KKSG31 & 12 38 33.7 & -10 29 25.0 & -12.3 & 9.3 & 13.5 & wide N \\
CGCG 014-054 & 12 31 3.8 & +1 40 33.0 & -14.4 & 9.6 & 13.2 & wide N \\
NGC3115 & 10 05 14.0 & -7 43 7.0 & -20.8 & 9.7 & 5.9 & medium N \\
NGC4517 & 12 32 45.5 & +0 06 54.0 & -20.2 & 9.7 & 7.3 & wide N \\
KKSG16 & 09 59 47.5 & -9 20 36.0 & -12.2 & 9.7 & 13.6 & medium N \\
KKSG18 & 10 05 41.6 & -7 58 53.0 & -16.6 & 9.7 & 10.1 & medium N \\
KDG065 & 10 05 34.4 & -7 44 57.0 & -13.6 & 9.7 & 12.2 & medium N \\
KKSG17 & 10 01 38.4 & -8 14 56.0 & -14.8 & 9.7 & 12.8 & medium N \\
MCG -01-26-009 & 10 01 33.6 & -6 31 30.0 & -14.0 & 9.7 & 13.6 & medium N \\
UGCA193 & 10 02 36.2 & -6 00 43.0 & -15.8 & 9.7 & 12.0 & medium N \\
KKSG15 & 09 55 10.5 & -6 16 12.0 & -15.0 & 9.7 & 12.6 & medium N \\
NGC0024 & 00 09 56.4 & -24 57 48.0 & -18.3 & 9.9 & 9.0 & medium S \\
ESO473-024 & 00 31 22.5 & -22 45 57.0 & -13.7 & 9.9 & 14.0 & medium S \\
\hline
\hline
\end{tabular}
\tablecomments{Data from \citet{karachentsev2013}. }
\end{center}

\label{tbl:nbgs3}
\end{table*}

%

\begin{figure*}[htbp]
\begin{center}
\includegraphics[width=\textwidth]{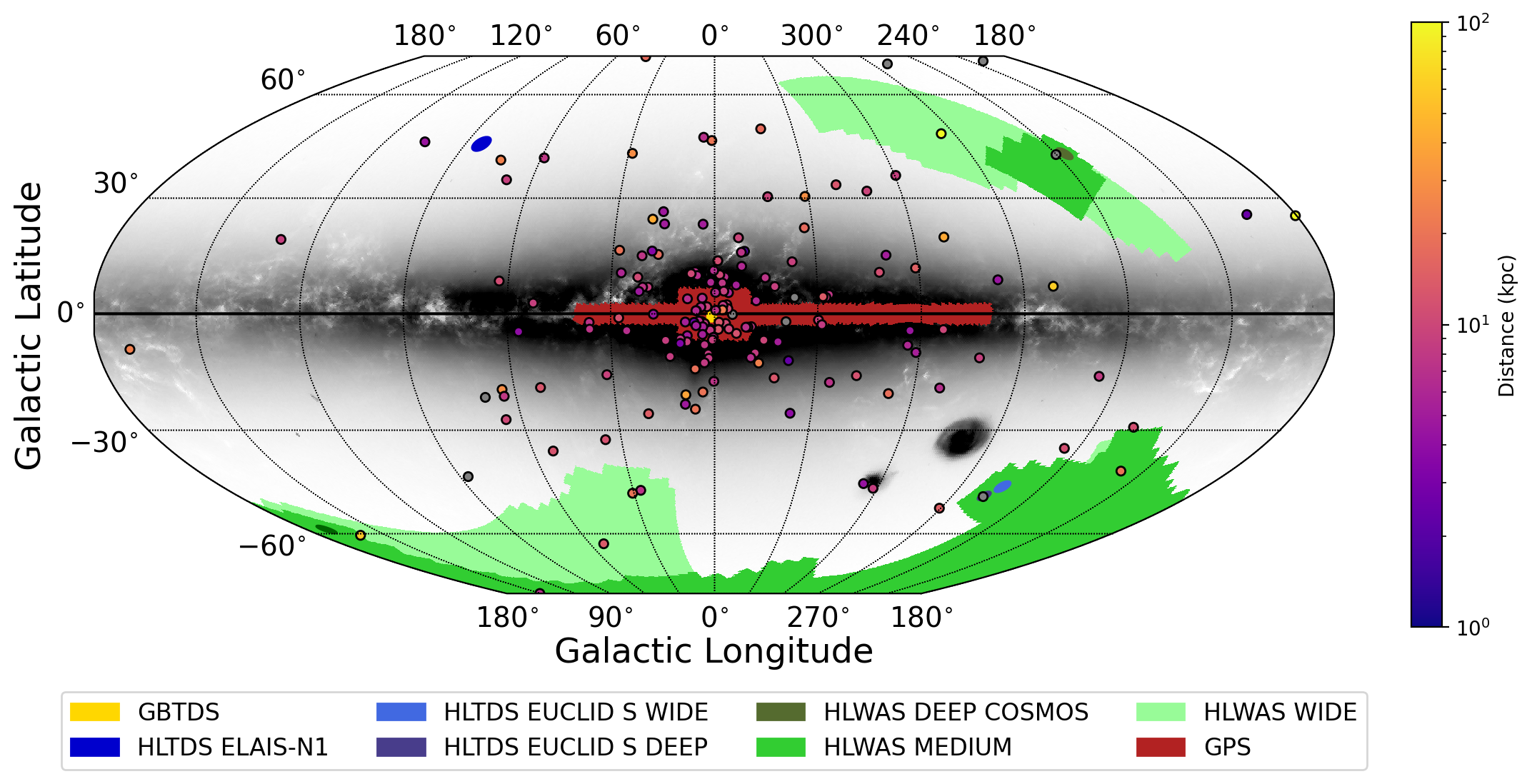}
\caption{Roman core community survey footprints and Milky Way globular clusters (colored by distance). Colors are the same as Figure~\ref{fig:streams-allsky}. Globular cluster positions and distances are from \citet{2025OJAp....8E.142P} and \citet{Baumgardt2021}.}
\label{fig:gcs-allsky}
\end{center}
\end{figure*}

\begin{figure*}[p]
    \begin{center}
        \includegraphics[width=\textwidth]{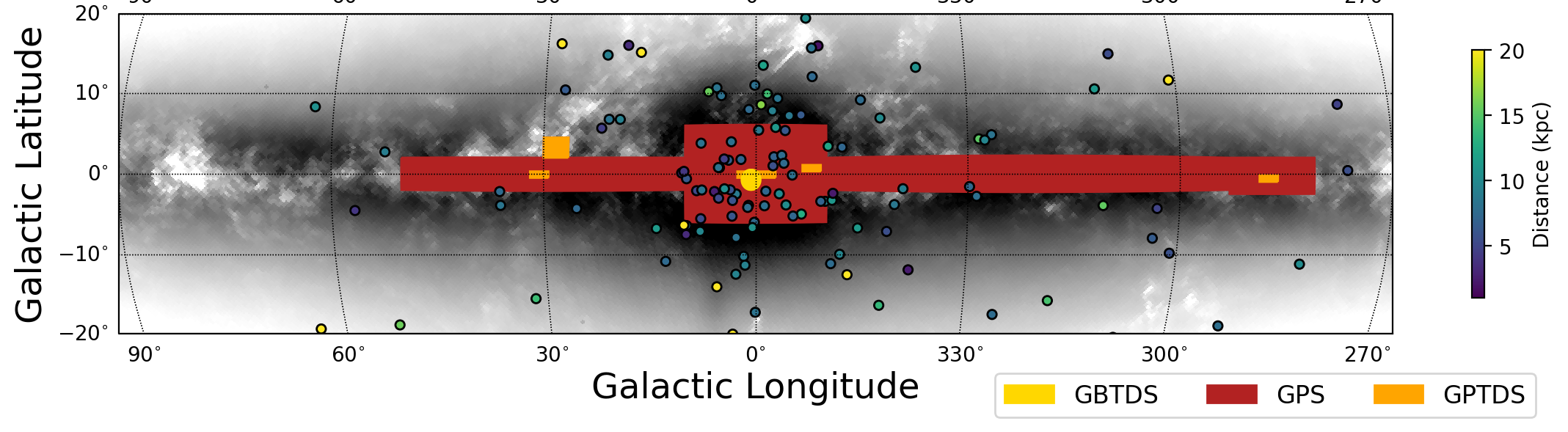} \\
        \includegraphics[width=0.45\textwidth]{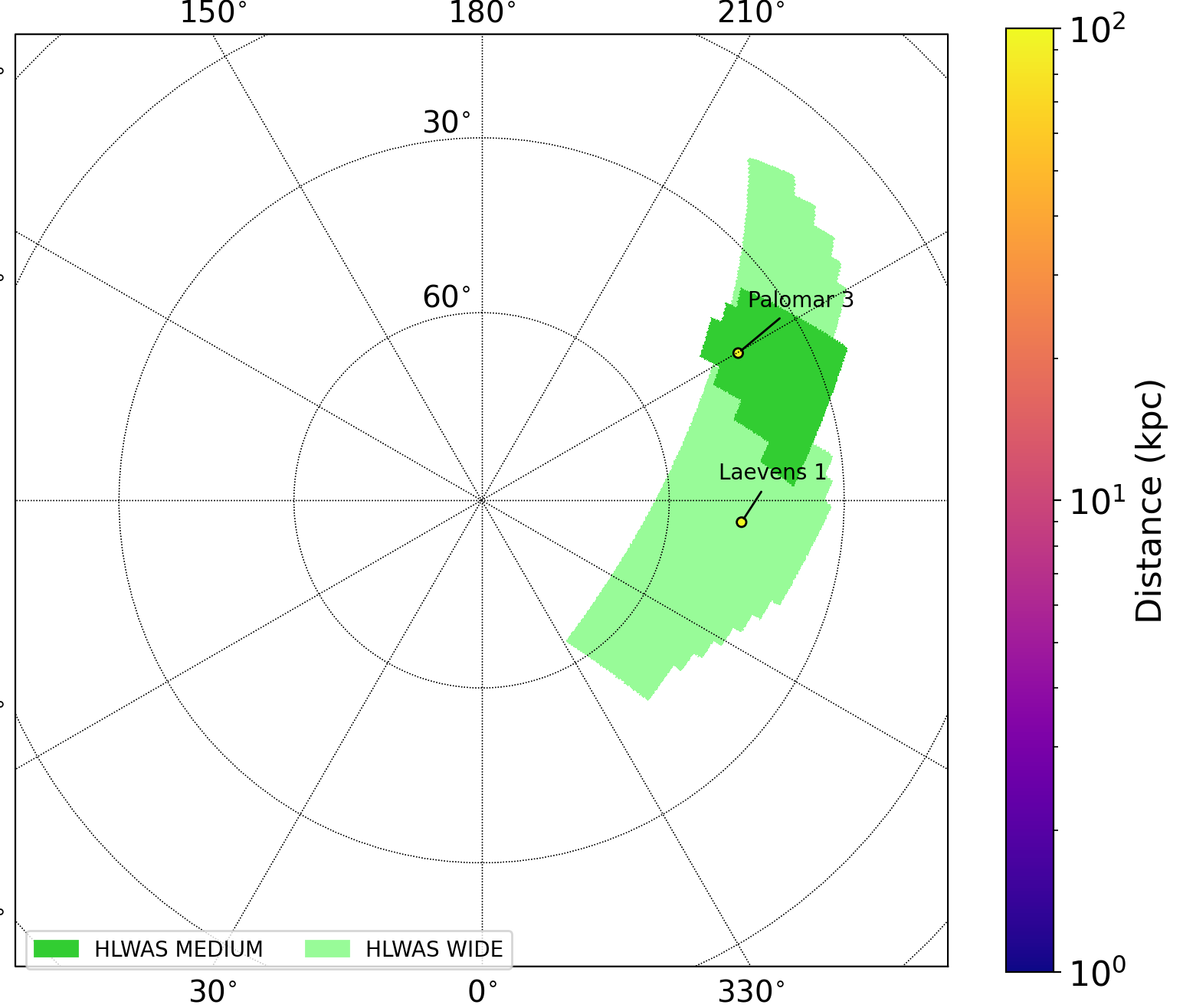}
        \includegraphics[width=0.45\textwidth]{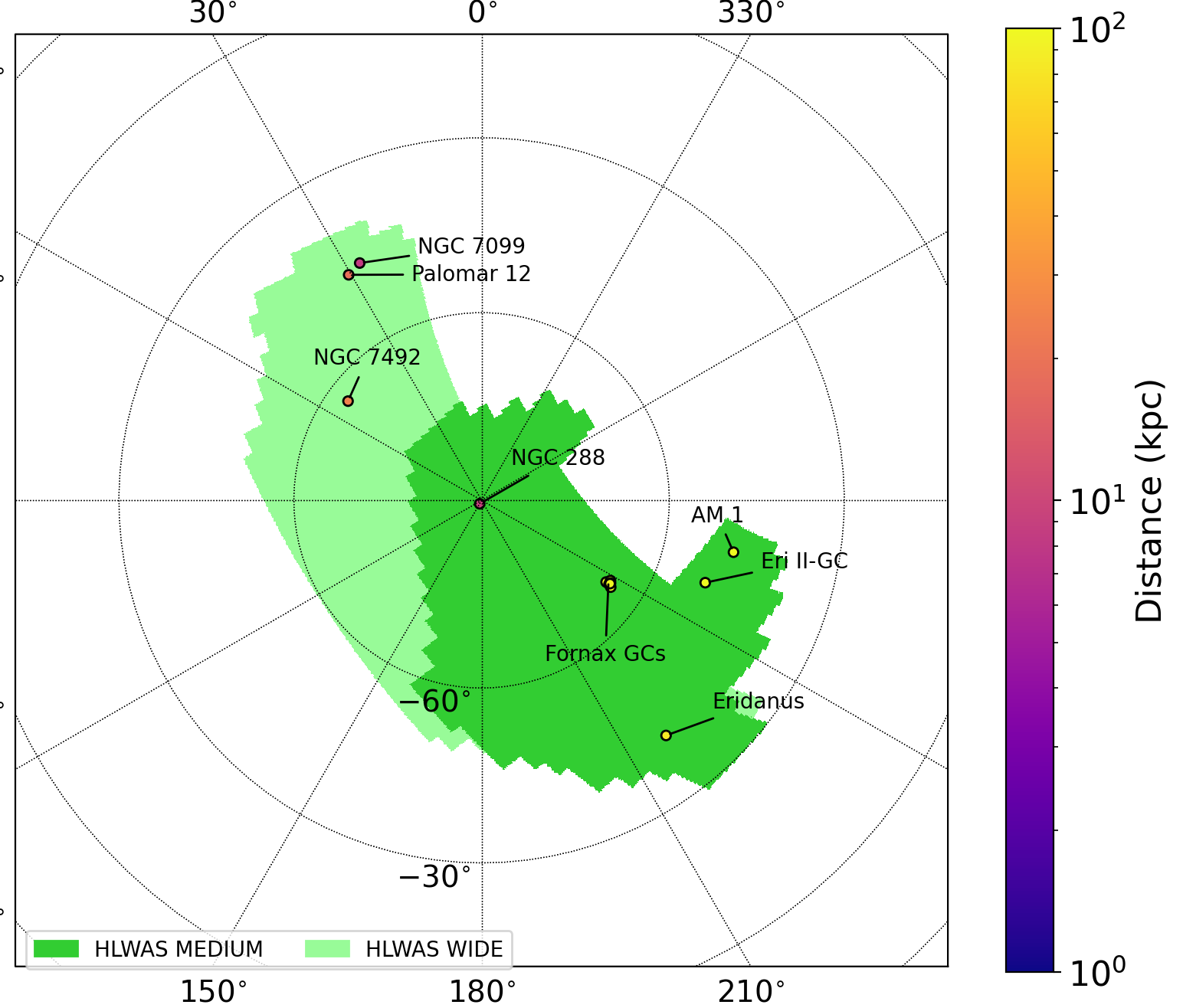}
        \caption{Globular clusters colored by distance in the Galactic Plane Survey (top), the HLWAS on the north Galactic cap (bottom left), and the HLWAS on the south Galactic cap (bottom right). Tables \ref{tbl:gcs_GPS} and \ref{tbl:gcs_HLWAS} list the clusters in the footprints of the GPS and HLWAS respectively. Globular cluster positions and distances are from \citet{2025OJAp....8E.142P} and \citet{Baumgardt2021}.}
        \label{fig:gcs-gps-hlwas}
    \end{center}
\end{figure*}

\begin{table*}
\caption{Globular clusters in the GPS}
\begin{center}
    \begin{tabular}{|lcccc|}
    \hline \hline
        Name & RA & Dec & $\ell$ & $b$ \\
        \hline
2MASS GC-01 & 18 08 21.8 & -19 49 47.0 & +10 28 15.6 & +00 06 0.3 \\
2MASS GC-02 & 18 09 36.5 & -20 46 44.0 & +09 46 55.6 & +00 36 54.6 \\
BH 261 & 18 14 6.6 & -28 38 6.0 & +03 21 42.1 & -05 16 13.4 \\
Djorgovski 1 & 17 47 28.7 & -33 03 59.0 & -03 19 30.1 & -02 29 0.9 \\
Djorgovski 2 & 18 01 49.1 & -27 49 32.9 & +02 45 48.6 & -02 30 29.9 \\
FSR 1735 & 16 52 10.6 & -47 03 29.0 & -20 48 44.5 & -01 51 11.6 \\
HP 1 & 17 31 5.2 & -29 58 54.0 & -02 34 29.2 & +02 06 54.1 \\
Liller 1 & 17 33 24.6 & -33 23 22.4 & -05 09 34.9 & +00 09 38.3 \\
NGC 6304 & 17 14 32.3 & -29 27 43.3 & -04 10 27.8 & +05 22 31.9 \\
NGC 6316 & 17 16 37.2 & -28 08 23.4 & -02 49 28.7 & +05 45 53.8 \\
NGC 6355 & 17 23 58.4 & -26 21 10.2 & +00 24 53.6 & +05 25 43.4 \\
NGC 6380 & 17 34 28.5 & -39 04 10.3 & -09 49 4.8 & -03 25 18.7 \\
NGC 6401 & 17 38 36.5 & -23 54 34.6 & +03 27 1.4 & +03 58 48.5 \\
NGC 6440 & 17 48 52.8 & -20 21 37.5 & +07 43 43.5 & +03 48 2.5 \\
NGC 6441 & 17 50 13.1 & -37 03 5.2 & -06 28 4.0 & -05 00 20.9 \\
NGC 6453 & 17 50 51.7 & -34 35 54.5 & -04 16 55.1 & -03 52 19.8 \\
NGC 6522 & 18 03 34.1 & -30 02 2.3 & +01 01 28.5 & -03 55 31.9 \\
NGC 6528 & 18 04 49.6 & -30 03 20.8 & +01 08 18.8 & -04 10 26.7 \\
NGC 6540 & 18 06 8.6 & -27 45 55.0 & +03 17 6.0 & -03 18 46.3 \\
NGC 6544 & 18 07 20.1 & -24 59 53.6 & +05 50 11.5 & -02 12 8.8 \\
NGC 6553 & 18 09 17.5 & -25 54 29.0 & +05 15 11.8 & -03 01 45.0 \\
NGC 6626 & 18 24 32.9 & -24 52 11.4 & +07 47 53.5 & -05 34 50.5 \\
Palomar 6 & 17 43 42.2 & -26 13 30.0 & +02 05 24.0 & +01 46 43.7 \\
Terzan 1 & 17 35 47.2 & -30 28 54.4 & -02 26 32.8 & +00 59 28.0 \\
Terzan 2 & 17 27 33.1 & -30 48 8.4 & -03 40 50.2 & +02 17 53.3 \\
Terzan 4 & 17 30 39.0 & -31 35 43.9 & -03 58 33.5 & +01 18 27.7 \\
Terzan 5 & 17 48 4.8 & -24 46 44.6 & +03 50 22.1 & +01 41 12.6 \\
Terzan 6 & 17 50 46.9 & -31 16 30.6 & -01 25 39.2 & -02 09 47.8 \\
Terzan 9 & 18 01 38.8 & -26 50 23.0 & +03 36 11.3 & -01 59 19.6 \\
Terzan 10 & 18 02 57.8 & -26 04 1.0 & +04 25 16.5 & -01 51 51.6 \\
Terzan 12 & 18 12 15.8 & -22 44 31.0 & +08 21 29.2 & -02 06 2.8 \\
Ton 2 & 17 36 10.1 & -38 33 22.0 & -09 12 23.9 & -03 25 25.0 \\
UKS 1 & 17 54 27.2 & -24 08 43.0 & +05 07 31.4 & +00 45 50.3 \\
FSR 1716 & 16 10 30.0 & -53 44 56.0 & -30 13 18.9 & -01 35 33.3 \\
FSR 1776 & 17 54 14.3 & -36 09 8.6 & -05 16 47.8 & -05 15 0.1 \\
ESO 456-29 & 17 58 36.2 & -32 01 12.0 & -01 13 58.6 & -03 58 36.0 \\
Gran 5 & 17 48 54.7 & -24 10 12.0 & +04 27 33.1 & +01 50 18.8 \\
Kronberger 49 & 18 10 24.0 & -23 20 24.0 & +07 37 37.8 & -02 00 44.1 \\
VVV CL0001 & 17 54 42.5 & -24 00 53.0 & +05 16 2.9 & +00 46 47.1 \\
VVV CL160 & 18 06 57.1 & -20 00 54.0 & +10 08 52.0 & +00 17 59.5 \\
VVV CL002 & 17 41 6.2 & -28 50 42.0 & +00 26 29.3 & +00 53 20.8 \\
Mercer 5 & 18 23 19.7 & -13 40 8.4 & +17 35 35.9 & +00 06 42.8 \\
Ryu 879 & 18 45 28.1 & -5 11 31.2 & +27 37 52.7 & -01 02 29.7 \\
    \hline \hline
    \end{tabular}
    \end{center}
    \label{tbl:gcs_GPS}
\end{table*}

\begin{table*}
\caption{Globular clusters in the HLWAS}
\begin{center}
    \begin{tabular}{|lccccl|}
    \hline \hline
        Name & RA & Dec & $\ell$ & $b$ & Survey\\
        \hline
        Palomar 3 & 10 05 31.6 & +0 04 18.0 & -119 51 34.6 & +41 51 48.9 & medium N \\
        Laevens 1 & 11 36 16.0 & -10 52 37.9 & -85 11 34.7 & +47 50 50.5 & wide N \\
        AM 1 & 03 55 2.3 & -49 36 55.0 & -101 38 19.2 & -48 28 14.6 & medium S \\
        Eridanus & 04 24 44.5 & -21 11 12.4 & -141 53 38.2 & -41 19 55.3 & medium S \\
        NGC 288 & 00 52 45.2 & -26 34 57.4 & +151 17 6.6 & -89 22 49.5 & medium S \\
        NGC 7099 & 21 40 22.1 & -23 10 47.5 & +27 10 44.9 & -46 50 7.7 & wide S \\
        NGC 7492 & 23 08 26.7 & -15 36 41.3 & +53 23 10.8 & -63 28 39.5 & wide S \\
        Palomar 12 & 21 46 38.8 & -21 15 9.4 & +30 30 36.4 & -47 40 53.9 & wide S \\
        Eri II-GC & 03 44 22.4 & -43 32 0.1 & -110 13 9.4 & -51 38 32.8 & medium S \\
        Fornax-GC1 & 02 37 1.9 & -34 11 1.0 & -123 16 31.9 & -66 17 56.7 & medium S \\
        Fornax-GC2 & 02 38 44.1 & -34 48 30.0 & -121 55 17.0 & -65 50 20.1 & medium S \\
        Fornax-GC3 & 02 39 48.1 & -34 15 30.0 & -123 20 12.4 & -65 43 19.8 & medium S \\
        Fornax-GC4 & 02 40 7.6 & -34 32 10.0 & -122 42 3.3 & -65 36 29.3 & medium S \\
        Fornax-GC5 & 02 42 21.1 & -34 06 7.0 & -123 54 45.6 & -65 13 36.2 & medium S \\
        Fornax-GC6 & 02 40 6.9 & -34 25 19.2 & -122 58 19.4 & -65 37 49.8 & medium S \\
        \hline
\end{tabular}
\end{center}
    \label{tbl:gcs_HLWAS}

\end{table*}

\section{Milky Way Clusters}
\label{sec:gcs}

The majority of the Milky Way's known globular clusters are near the Galactic Center (Figure \ref{fig:gcs-allsky}, so the bulk of those observed by Roman's wide-field surveys are in the sky area covered by the Galactic Plane Survey (GPS). Using the survey definition submitted to the mission (\citealt{2025arXiv251107494G}; B. Benjamin, priv. comm) shows that 37 of the known GCs from \citet{Baumgardt2021} are in the footprint of the GPS (Figure \ref{fig:gcs-gps-hlwas} top; Table \ref{tbl:gcs_GPS}). The observations of the GPS fields are optimized for proper motions, and the long time-baselines for clusters with HST observations will further improve the measured phase-space positions of these clusters, and perhaps for some even measure their internal motions as for nearby satellite galaxies (Figure \ref{fig:HLWAS_Medium+HST_astrometry_errs}). 

A few GCs also fall in the footprint of the HLWAS: two in the Galactic north and seven in the Galactic south (Figure \ref{fig:gcs-gps-hlwas} bottom; Table \ref{tbl:gcs_HLWAS}). The GCs in the medium field, in particular, will be observed with three filters over the first two years of observations. 

Roman's wide field of view offers particular advantages for GC studies. The broad coverage will permit systematic searches for extratidal features \citep[e.g.][]{2025arXiv251014924C} and map the outer density profiles \citep[e.g.][]{2012MNRAS.419...14C} around all the GCs observed by Roman. It will also allow for cluster-wide studies of the multiple populations observed in small fields by HST \citep[e.g.][]{2019A&ARv..27....8G}. Proper motions of stars in multiple populations in the outskirts of GCs are particularly constraining for theories of their formation, since these stars have the longest dynamical times (e.g. \citealt{2008MNRAS.391..825D}). Roman's infrared sensitivity, combined with high-quality proper motions to select cluster members, also offers the promise of mapping GC stellar populations below the hydrogen-burning limit, allowing the study of brown dwarfs in GCs \citep{2016ApJ...817...48D,astrometry2019}---a regime in metallicity and age that only now is beginning to be explored \citep[e.g.][]{nardiello2023,Gerasimov_2024,2024A&A...689A..59S,2025A&A...701A.169S,2025A&A...694A..68S,2025AN....34670042B}. These studies as well will benefit from Roman's ability to map the lower-density outskirts of clusters, where contamination from the light of nearby RGB stars is negligible.

\section{Summary}
The wide-field surveys planned for Roman will provide a rich new data source for studies of the Milky Way and its neighbors, including nearly 60 known streams (Section \ref{sec:streams}), 39 nearby galaxies including 14 Milky Way satellites (Section \ref{sec:nbgs}), and 46 Milky Way globular clusters (Section \ref{sec:gcs}). For many of these systems, Roman has the potential to extend the proper motions measured by the Gaia mission to far fainter magnitudes (Section \ref{sec:roman_PMs}), although  in the current HLWAS scheduling plan this is largely a missed opportunity. A second epoch, either in the past with HST or in a future Roman program, is needed to allow Roman to reach its full astrometric potential. Nevertheless, the depth of the planned surveys and their unprecedented image quality over large sky area at infrared wavelengths will be powerful in the near field.

To aid in planning and analysis, the data and code used to generate the figures and tables in this paper are available publicly at \url{github.com/Dynamics-Penn/roman-nearfield} \citep{robyn_sanderson_2026_18961307}. The footprints for the core community surveys are also available as part of the \texttt{skyproj} package \citep[\url{skyproj.readthedocs.io/en/latest/surveys.html\#romanhlwasskyproj},][]{skyproj_citation}. We expect to update both sources with any changes to the footprints (although none are anticipated at this late stage).

\begin{acknowledgments}
RES, BW, and ACRT acknowledge support from NASA Award 80NSSC24K0084. RES and BW were also supported by NASA ROSES grant 22-ROMAN11-0011, contract number 80NM0024F0012, via a JPL subaward. RES and BW are grateful to the members of the HLIS Cosmology PIT, especially David Weinberg, Chris Hirata, and Kaili Cao, for in-depth discussions of HLWAS scheduling. RES thanks Bob Benjamin for providing the updated footprint for the GPS. SWJ gratefully acknowledges support from a Guggenheim Fellowship.
K.A.M. acknowledges support from the University of Toronto’s Eric and Wendy Schmidt AI in Science Post-doctoral Fellowship, a program of Schmidt Sciences. AJL is supported by NASA through the NASA Hubble Fellowship grant \#HST-HF2-51580.001 awarded by the Space Telescope Science Institute, which is operated by the Association of Universities for Research in Astronomy, Inc., for NASA, under contract NAS5-26555. SP was supported by a research grant (VIL53081) from VILLUM FONDEN. This work was co-funded by the European Union (ERC, BeyondSTREAMS, 101115754). Views and opinions expressed are, however, those of the author(s) only and do not necessarily reflect those of the European Union or the European Research Council. Neither the European Union nor the granting authority can be held responsible for them. 

\end{acknowledgments}

\vspace{5mm}
\facilities{\textit{Roman}, \textit{Gaia}}

\software{\texttt{astropy} \citep{astropy_2013,astropy_2018,astropy_2022}, \texttt{BP3M} \citep{McKinnon_2024}, \texttt{Gaia+Roman Astrometry Simulation Tool} \citep{McKinnon_2026}, \texttt{jupyter} \citep{jupyter_citation}, \texttt{matplotlib}  \citep{matplotlib_citation}, \texttt{numpy} \citep{numpy_citation}, \texttt{pandas} \citep{pandas_citation_2010,pandas_citation_2020}, \texttt{pandeia} \citep{Pontoppidan_2016}, \texttt{roman-nearfield} \citep{robyn_sanderson_2026_18961307}, \texttt{scipy} \citep{scipy_citation}, \texttt{stpsf} \citep{Perrin_2012,Perrin_2025}}, \texttt{healpy} \citep{healpy_citation}, \texttt{shapely} \citep{shapely_citation}, \texttt{galstreams} \citep{Mateu2023}, \texttt{skyproj} \citep{skyproj_citation}

\bibliography{gsrefs}{}
\bibliographystyle{aasjournalv7}

\end{document}